# THE EVOLUTION OF CIRCUMSTELLAR DISKS IN OPHIUCHUS BINARIES

J. Patience[1,2], R. L. Akeson[3], E. L. N. Jensen[4]
[1] School of Physics, University of Exeter, Stocker Road, Exeter, EX4 4QL, Devon, United Kingdom; patience@astro.ex.ac.uk
[2] California Institute of Technology, Department of Astronomy, MS 105-24, 1200 E. California Blvd., Pasadena, CA 91125
[3] Michelson Science Center, California Institute of Technology, MS 100-22, 1201 E. California Blvd., Pasadena, CA 91125; rla@ipac.caltech.edu
[4] Swarthmore College, Swarthmore, PA 19081; ejensen1@swarthmore.edu

**Abstract**
Four Ophiuchus binaries, two Class I systems and two Class II systems, with separations of ~450 – 1100 AU were observed with the Owens Valley Radio Observatory (OVRO) millimeter interferometer. In each system, the 3mm continuum maps show dust emission at the location of the primary star, but no emission at the position of the secondary. This result is different from observations of less evolved Class 0 binaries in which dust emission is detected from both sources. The nondetection of secondary disks is, however, similar to the dust distribution seen in wide Class II Taurus binaries. The combined OVRO results from the Ophiuchus and Taurus binaries suggest that secondary disk masses are significantly lower than primary disk masses by the Class II stage, with initial evidence that massive secondary disks are reduced by the Class I stage. Although some of the secondaries retain hot inner disk material, the early dissipation of massive outer disks may negatively impact planet formation around secondary stars. Masses for the circumprimary disks are within the range of masses measured for disks around single T Tauri stars and, in some cases, larger than the Minimum Mass Solar Nebula. More massive primary disks are predicted by several formation models and are broadly consistent with the observations. Combining the 3mm data with previous 1.3mm observations, the dust opacity power law index for each primary disk is estimated. The opacity index values are all less than the scaling for interstellar dust, possibly indicating grain growth within the circumprimary disks.

## 1. Introduction

Comprehensive surveys of main sequence binaries with 0.03 AU to $8 \times 10^4$ AU orbits (Duquennoy & Mayor 1991) show a total fraction of binaries exceeding 50%, and high resolution imaging and spectroscopic surveys of nearby star-forming regions reveal an even higher proportion of binary and multiple systems (Ghez *et al.* 1993, Leinert *et al.* 1993, Simon *et al.* 1995, Mathieu *et al.* 2000, Haisch *et al.* 2004, Duchene *et al.* 2004). These studies emphasize the importance of considering the environment of binary stars for a comprehensive understanding of the star and planet formation process. By observing young binary systems, it is possible to investigate the early conditions of star and planet formation. In particular, circumstellar disks are still present in the pre-main sequence stage. These critical structures maintain a reservoir of material for planet formation and provide a means of accreting mass from the envelopes surrounding the youngest stars and onto the stellar surface throughout the pre-main sequence evolution.

This study of young binaries at different evolutionary stages in the Ophiuchus star-forming region is designed to probe the cooler outer disk material that comprises most of the disk mass. Our sample includes targets at two different evolutionary stages. Young stars are grouped into different evolutionary classes (I, II, III) based on the shape of their spectral energy distributions defined by the spectral index $a = d\log(\lambda F_\lambda)/d\log(\lambda)$, with decreasing values of $a$ from Class I to III (Lada & Wilking

1984). The binaries selected for this survey cover the Class I and Class II stages. Combined with previous data on Class 0 Ophiuchus binaries (Wootten 1989, Looney *et al*. 2000) the new results measure the evolution of circumstellar disks throughout the pre-main sequence lifetime of binaries separated by several hundred AU in this star-forming region.

The sample of Ophiuchus binaries is defined in Sect. 2, followed by a description of the observations and data analysis in Sect. 3. The observations include both interferometric millimeter data and optical spectra. Sect. 4 reports the results of the mm continuum and spectral line observations and the analysis of the high resolution optical spectra. While the millimeter data probe the bulk of the disk material at cooler temperatures, the optical spectra provide Hα equivalent widths for some of the stars, an indicator of accretion activity near the stellar photosphere. The discussion, Sect. 5, addresses the physical parameters of the disks such as mass and dust opacity, and compares these values with previous observations and theoretical models. Finally, Sect. 6 provides a summary.

**2. The Sample**

To investigate the evolution of disk material in binary systems, a sample of Class I and Class II Ophiuchus binaries was selected for observation with the Owens Valley millimeter-wave array (OVRO). Class I binaries were drawn from a NIR/mid-IR multiplicity survey of ρ Oph and Serpens clouds (Haisch *et al*. 2002), and Class II binaries were taken from extensive IR speckle/imaging and CCD imaging surveys of Southern T Tauri stars (Simon *et al*. 1995, Reipurth & Zinnecker 1993). The targets satisfied two main selection criteria: millimeter fluxes sufficient for detection and separations in the 100's – 1000 AU range. The current sample is small and this limits the statistical significance; however, observations of Class I binaries in Taurus are being conducted.

Of the four systems targeted for this project IRS 43 and L1689SNO2 are Class I and SR 24 and Elias 30 are Class II. Basic information about each binary is listed in Table 1: evolutionary class, coordinates, binary parameters, spectral types, alternate names, and the spectral types and mass ratios of the primary and secondary stars. The evolutionary class of these sources has been determined from both 2.2-25 µm (Wilking *et al*. 1989) and 2-10 µm (Greene *et al*. 1994, Luhman & Rieke 1999) measurements; while the exact value of the spectral energy distribution slope differs somewhat between studies, the Class I targets always have flat or rising distributions and the Class II targets always have declining fluxes towards longer wavelengths. Spectral types are available for three of the four primaries, but only the Class II secondaries (Luhman & Rieke 1999, Greene & Lada 2002, Prato *et al*. 2003). In cases where different spectral types have been measured, the estimate from higher resolution spectroscopy is used.

The range of binary separations was chosen to be larger than individual disks. Based on the $5 \times 10^{-5}$ stars/square arcsecond density of field stars brighter than K=12mag in the Ophiuchus region (Greene & Young 1992), the probability that one of the systems is a chance projection rather than a physically associated pair is less than 3%. Additionally, the ~100's – 1000 AU range is similar to previous mm studies of T Tauri binaries and ensured that each system was resolvable with OVRO. The expected beam size was 3".8 x 2".5, so the minimum separation considered was approximately 3 arcseconds, or 480 AU, assuming a distance of 160 pc. Among the seven Ophiuchus binaries in the ~3-10" range, only four have the required flux level, estimated by scaling the 1.3mm flux from an IRAM 30m survey of Ophiuchus (André & Montmerle 1994). For two of the systems, SR 24 and L1689 SNO2, the secondaries are actually subarcsecond pairs, unresolvable with OVRO (Simon et al. 1995, Ratzka et al. 2005).

For simplest comparison with previous work, the distance to Ophiuchus is taken as the traditional value of 160 pc (Chini 1981). As detailed in Bontemps *et al*. (2001), *Hipparcos* measurements of the distance to Upper Sco OB association (de Zeeuw *et al*. 1999) coupled with studies of the geometry of the Ophiuchus members (de Geus *et al*. 1989) suggest that a revised distance of 140 ± 10 pc may be more accurate (see also Mamajek 2007). If the closer distance is used, then the inferred disk masses need to be lowered by 23%; no other quantity reported in this paper is affected.

**3. Observations and Data Analysis**
**3.1 OVRO mm array data**

Observations for each source were obtained with the six-antenna OVRO millimeter-wave array during February, March, and April of 2003. The central frequency of the observations was 112 GHz, and the total continuum bandwidth was 4 GHz. Three spectral lines CO(1-0) (115.2712 GHz), $^{13}$CO(1-0) (110.20135 GHz), and $C^{18}O$(1-0) (109.78216 GHz) were observed simultaneously, providing information on both optically thin and thick transition lines. For CO(1-0) and $^{13}$CO(1-0), the velocity resolution was 0.22 km/s over a 20 km/s range, while the weaker line $C^{18}O$(1-0) was observed with coarser resolution of 0.65 km/s over a 20 km/s range.

The observations were made in two separate configurations of the array with baselines ranging from 15 to 220 m, which translate into angular scales from 2".5 to 37" (e.g. Westpfahl 1999). To produce similar spatial frequency coverage on each target, two of the binaries and the gain calibrator J1624-354 were observed during each four-hour sequence. An internal noise source and the bright quasar 3C273, 3C345, or 3C454.3 were used to calibrate the spectral passband, and observations of either the planet Uranus or Neptune provided an absolute flux standard.

Each four-hour observing track included a pair of Ophiuchus sources, either the Class I targets IRS43 and L1689SNO2 or the Class II targets SR24S and Elias30. For these low declination sources, the system temperatures at zenith ranged from ~400 K to 600 K and the amount of precipitable water vapor ranged from 2.9 mm to 6.5 mm. The rms noise level in the tracks used for mapping was typically ~3 mJy beam$^{-1}$.

Initial data reduction steps were performed with MMA software (Scoville et al. 1993): flux calibration with the planet observations, gain calibration of the recorded amplitudes and phases to account for atmospheric fluctuations, data editing, and passband calibration of the spectral data with the internal noise source and a bright quasar. Subsequent analysis of the reduced data was conducted using the MIRIAD software (Sault *et al*. 1995) tasks to generate CLEANed maps of each source from which source sizes and fluxes were derived. The final maps use a weighting parameter value of 2.0 that provides the best signal-to-noise ratio, with the tradeoff of higher sidelobes and a larger beam.

For the line data, a series of channel maps for CO(1-0), $^{13}$CO(1-0), and $C^{18}O$(1-0) were made for each source with MIRIAD tasks. After the data from several tracks were averaged and outlying points deleted, the individual channels separated by 0.22km/s (CO(1-0) and $^{13}$CO(1-0)) or 0.65km/s ($C^{18}O$(1-0)) were mapped. For increased signal-to-noise, the CO(1-0) and $^{13}$CO(1-0) channels were also integrated in steps of 2 to 16 channels (or 0.4 to 3.5 km/s) to search for line emission. Similarly integrated $C^{18}O$(1-0) maps were also generated.

**3.2 CTIO spectroscopy**

Three of the four binary systems were also observed with the echelle spectrograph on the CTIO Blanco 4-meter telescope on 2002 August 4 (SR 24), and 2003 April 12 and 14 (Elias 30 and L1689SNO2). The spectrograph setup gave wavelength coverage of approximately 5,000--8,000 Å with 0.08 Å/pixel. With a one-arcsecond slit width, this gave a spectral resolution (as measured from narrow lines in a ThAr comparison lamp spectrum) of *R = 40,000* at the Hα line.

The wide pairs in all three binary systems were resolved, yielding separate spectra of the primary and secondary stars. For the hierarchical triples SR 24 and L1689 SNO2, the spectra of the secondary components included both stars in the close pairs. Exposure times were 30 minutes per star for all objects except for L1689 SNO 2, for which the total exposure time was 60 minutes per star. The data were reduced using standard routines for echelle spectra in *IRAF*. The instrumental setup used did not allow adequate room between adjacent echelle orders on the chip to extract a good sky spectrum.

## 4. Results
### 4.1 3mm Continuum

The 3mm OVRO maps for the four binary systems are shown in Figures 1a-d as contour maps overlaid on 2MASS $K_s$ images which are centered on the primary of each pair. The alignment of the two datasets is based on the absolute positions. The 2MASS coordinate error ellipses for these sources are at most 0".09 (http://irsa.ipac.caltech.edu/applications/Gator). The uncertainties of the OVRO positions can be estimated either by dividing the beam size by the signal-to-noise ratio of the peak of the detection in the final map (Fomalont 1999) or by dividing the beam size by a factor dependent upon the angular distance from target to calibrator and the baseline solution uncertainty (Morita 1992). Both methods give values of about 0".1 -- much less than the binary separations -- so there is no ambiguity in assigning the flux to the correct component.

The coordinates of the mm peak in each image are given in Table 2 along with the beam sizes and positional uncertainties. These coordinates and sizes were determined by fitting a Gaussian convolved with the beam to the mm detection. Two of the sources – IRS 43 and SR 24 – are slightly resolved, while the other two targets are point sources. In all cases, only the primary component has detectable mm continuum emission. As the likely origin of the millimeter emission is a disk, this suggests that circumsecondary disks are significantly depleted even at the earlier evolutionary stage of Class I sources. Nevertheless, the optical spectra (see Section 4.3) demonstrate that at least one secondary source (SR 24N) retains hot inner disk material responsible for its H$\alpha$ accretion signature.

The continuum fluxes for the primaries were calculated from the final maps by summing the flux in a region centered on the primary star position; the boundary size was set to include contours corresponding to 2$\sigma$ and above. When possible, the 3mm flux has been corrected for a free-free emission component based on published cm flux measurements (Girart et al. 2004) extrapolated to 3mm. Upper limits for the secondaries were estimated as three times the rms noise in the map. The flux uncertainty includes the rms noise in the map and the absolute flux calibration (11% based on the standard deviation of the flux measurements of the gain calibrator); they are listed separately in Table 3 and combined in quadrature in Table 4. All of the 3mm continuum fluxes and detection limits, along with the sizes of the two resolved sources, are listed in Table 3.

### 4.2 Spectral Line Images

Of the four sources observed, only IRS 43 had significant emission detected in each spectral line. SR 24 showed emission in CO(1-0), but not the other lines. L1689SNO2 and Elias 30 were not detected, although all targets had similar sensitivity limits.

For IRS 43, Figures 2a-c show the $^{13}$CO(1-0), CO(1-0), and C$^{18}$O(1-0) images, respectively. The map of the $^{13}$CO(1-0) in Figure 2a reveals largely symmetric red and blue shifted emission emanating from the primary star, oriented in the East-West direction. Red-shifted emission covers the velocity range from 5.0-6.6km/s and the blue-shifted emission covers velocities from 0.96-2.5km/s away from the system velocity of ~4km/s. The CO(1-0) (Figure 2b) displays red and blue-shifted emission over similar velocity ranges, though the CO(1-0) distribution is more compact than that of $^{13}$CO(1-0). The weaker detection of $C^{18}O(1-0)$ is plotted in Figure 2c and shows a distribution of gas on either side of the IRS 43 primary with velocities and locations similar to the outer parts of the $^{13}CO(1-0)$ data. Figure 3 shows the CO(1-0) channel maps covering both the blue and red lobes, and the higher velocity gas is located closer to the star than the lower velocity gas; this type of position-velocity relation is common for rotating disks or envelopes, but not outflows that have accelerated gas farther from the central star. Thus, the OVRO data seem more consistent with a disk or torus rather than an outflow. The greater spatial extent of the $^{13}CO(1-0)$ and $C^{18}O(1-0)$ is, however, unexpected for either a disk (e.g. Corder et al. 2005) or an outflow (Arce & Sargent 2006). If the $C^{18}O(1-0)$ emission originates in a disk-like structure then it may reveal the edge of a ring or torus.

Previous observations of IRS 43 present a complicated and, for some aspects, contradictory description of the circumstellar environment. Interpreting the OVRO gas line emission data as arising from a rotating disk or torus/flattened envelope is consistent with the majority of the extant data. NICMOS images show evidence of scattered light from the upper and lower surfaces of a disk oriented E-W (Terebey et al. 2001), with the dark lane sharing the same direction as the OVRO-detected CO emission. Figure 4 shows a composite of the NICMOS image and the CO(1-0) data and provides the strongest complementary set of data suggesting the IRS 43 3mm line emission originates in a rotating structure. Larger field infrared $H_2$ $v=1-0$ S(1) images of embedded HH objects in Ophiuchus identify emission knots that point back to the location of IRS 43 and have an angle of 22° East of North projected onto the plane of the sky (Grosso *et al.* 2001). An outflow mainly directed N-S is consistent with an E-W disk structure. Similar to the HH object orientation, VLA 3.6 cm and 6 cm observations of IRS 43 also show an extended structure with a position angle of 24° (Girart *et al.* 2000, Girart *et al.* 2004) at a separation of 0".6 from the 2MASS position of IRS 43. The VLA data have been interpreted as either an outflow from a second protostar that is more massive, deeply embedded, and at an earlier evolutionary state (Girart *et al.* 2004) or a jet from a single source since there is no indication of a second star at this location in the NICMOS images or NIRSPEC spectra (Greene & Lada 2002). Finally, 1.3mm CO(2-1) single dish maps previously obtained (Bontemps *et al.* 1996) reveal overlapping red and blue shifted lobes which were interpreted as a nearly pole-on outflow. The OVRO CO(1-0) map also shows overlapping red and blue shifted emission like the CO(2-1) data, but the velocity field is atypical of outflows, and a nearly edge-on disk suggested by NICMOS is difficult to reconcile with a pole-on outflow. In summary, for IRS 43 we find the OVRO CO data most consistent with originating from a disk.

In addition to the primary of IRS 43, the secondary in the Class II SR 24 system shows a weak CO(1-0) detection as indicated in Figure 5. Although there is no detection of a massive dust disk around the secondary based on the continuum upper limit, there appears to be a gas disk, a unique combination. The flux in the integrated channel map is 853 mJy which translates into a CO gas mass lower limit of $\geq 3.2 \times 10^{-5}$ Msun assuming a temperature of 30 K and using the method outlined in Scoville *et al.* (1986). This gas disk was also observed at CO(2-1) with the Submillimeter Array (Andrews & Williams 2005) and the higher signal-to-noise SMA map resolved the emission, indicating the material surrounds both stars in the close northern pair in a circumbinary ring. The SMA CO(2-1) data also detect the primary, but at a lower level which is below the threshold of the OVRO CO(1-0) map. The OVRO lower limit is consistent with, though less restrictive than, the lower limit of $\geq 3.3 \times 10^{-4}$ Msun obtained from the SMA data. The conversion to a gas mass requires an assumption that the gas is optically thin and this condition is unlikely to be realized for either the OVRO CO(1-0) or the SMA CO(2-1) data, making the calculated mass a lower limit.

### 4.3 Optical Spectra

Of the stars for which we have optical spectra (SR24, Elias 30, and L1689SNO2), the two components of SR 24 and the primary of Elias 30 show the strongest detections and are bright enough that the influence of sky and background emission is negligible. Thus, we discuss these spectra first.

Both SR 24 S and the unresolved binary SR 24 N show clear Li absorption and other photospheric features. Since their spectra have been discussed previously in the literature (e.g., Cohen & Kuhi 1979), we do not show the spectra nor discuss them further here except to focus on the Hα emission as a probe of possible accretion from circumstellar material. In both components the Hα emission (Figure 6) is broad and shows the double-peaked or centrally-absorbed structure commonly seen in classical T Tauri stars (e.g., Reipurth, Pedrosa, & Lago 1996). The emission equivalent width is 60 Å (for the southern primary SR 24 S) and 90 Å (for the unresolved northern binary SR 24 N); the full widths at 10% intensity are 600 km s$^{-1}$ and 560 km s$^{-1}$, respectively. Both the equivalent width and velocity width indicate strong accretion onto the stellar surfaces (White & Basri 2003), and Natta, Testi, & Randich (2006) derive accretion rates of $10^{-7}$ $M_{Sun}$ yr$^{-1}$ for both components based on their Pa β emission. This high accretion rate is especially interesting in the case of SR 24 N, since the northern binary does not have a massive dust disk visible in the continuum data, but only retains a gas disk detected in CO(1-0) emission (Section 4.2, and Andrews & Williams 2005).

In the Elias 30 system, the primary star Elias 30A shows weak but broad Hα emission (Figure 6). To our knowledge, this is the first detection of Hα emission in this source, though it has been known as an infrared excess source for decades. The emission equivalent width is 1.2 Å and the full width at 10% intensity is 450 km s$^{-1}$. This velocity width may be evidence for accretion (White & Basri 2003), but the line is quite weak and its equivalent width is well below the canonical lower limit of 5-10 Å for being considered a classical T Tauri star. Natta et al. (2006) did not detect Pa β emission from Elias 30A, and they set an upper limit on the accretion rate of $< 10^{-8.8}$ $M_{Sun}$ yr$^{-1}$. Li is seen in absorption with an equivalent width of 160 +/- 20 mÅ, relatively weak for a T Tauri star if its spectral type is G2.5 as found by Prato *et al.* (2003). Since the strength of the Li line declines with increasing effective temperature, the observed weakness of the line may indicate that the somewhat hotter spectral type of F4 found by Luhman & Rieke (1999) is more appropriate for Elias 30A. However, the extreme weakness of the Ca line at 6718 Å is puzzling, since F and G stars typically have a much stronger Ca line than that seen in Elias 30A (equivalent width ~ 50 mÅ, while our spectra of an F5V standard star at the same spectral resolution show Ca 6718 Å with a strength of 90 mÅ). Spectral veiling from accretion would lower the measured equivalent widths of absorption lines, and indeed the difference in Ca line strengths noted here is consistent with the veiling parameter $r_K = 0.78$ +/- 0.54 measured by Prato et al. (2003) if the optical veiling is the same as that measured in the K band. However, given the relatively low S/N of our spectrum it is not possible to say conclusively how much veiling is present. Absorption lines seen throughout the spectrum are broad; artificially broadening an F or early G stellar spectrum observed with the same spectrograph setup yields a good fit to most of the observed lines with $v \sin i$ ~ 50-75 km s$^{-1}$.

The spectrum of Elias 30B is too faint to separate the stellar emission from the background unambiguously, and we do not discuss it further here.

The spectra of both binary components of L1689SNO2 are quite faint and likely contain some contamination from diffuse Galactic and geocoronal Hα emission. However, our spectra of the primary and secondary sources were taken in the same two-hour period and differ in position by only 3". Thus, both sample roughly the same night-sky emission and Galactic background, and any marked differences between them should indicate differences in intrinsic source properties in the two binary components.

Comparison of the two spectra shows that the primary source+background spectrum shows a significantly stronger and broader Hα emission feature than the secondary source+background spectrum, indicating that L1689SNO2A has some intrinsic Hα emission. The equivalent width of this feature is ~ 6 Å, its FWHM is ~ 100 km s$^{-1}$, and its full width at 10% intensity is ~ 250 km s$^{-1}$ (Figure 6). This 10% width is similar to that seen in other Class I sources (*e.g.* White & Hillenbrand 2004). If one takes the secondary spectrum to be largely background emission and subtracts it from the primary spectrum as an approximate way of removing the background, the equivalent width and velocity widths in the resulting difference spectrum are not substantially different from those cited above. We note that these values are very uncertain due to the low S/N and large uncertainty on the continuum level of the spectrum, but the detection of broad Hα emission from the primary is robust.

Thus, all three of the primary stars in our sample that are detectable at visible wavelengths show Hα emission as well as mm continuum emission, indicating that accretion from the massive circumprimary disks is ongoing (though the evidence for accretion in Elias 30A is the weakest of the three). Interestingly, the one secondary star (SR 24 N) whose Hα emission has been measured also shows strong evidence of ongoing accretion, indicating that small circumstellar accretion disks can remain even in the absence of the larger-scale, massive disks seen in the interferometric maps.

## 5. Discussion

For all binaries observed, only the primary has detectable millimeter emission. Given the compact nature of the millimeter sources, the position coincident with the primary, and the evidence for Keplerian motion in the molecular line data for one target, the millimeter emission is modeled as arising from a disk encircling the primary in this discussion. The disk masses, lifetimes, and grain properties are important parameters to gauge the viability of planet formation and to test numerical models of binary formation.

### 5.1 Dust Opacity Scaling

The opacity, $\kappa$, of the dust in the disk is parameterized as a power law function of frequency, $\nu$, with the relation

$$\kappa(\nu) = \kappa_o (\nu/\nu_o)^\beta. \tag{1}$$

The power law index, $\beta$, depends on the size, composition, and structure of the dust grains, making a measurement of $\beta$ diagnostic of dust properties. Interstellar dust grains have values of ~2 (Hildebrand 1983), while models of spherical grains predict 1.53 at a temperature of 100K (Pollack *et al.* 1994), a higher temperature than the expected for these disks. Lower values of $\beta$ are also found in Mie calculations involving a distribution of dust particle sizes from sub-micron to more than 1mm (Miyake & Nakagawa 1993) rather than using submicron-sized interstellar dust (Draine & Lee 1984). The opacity power law index is also impacted if the grains are conducting material (Wright 1982) such as graphite needles, small amorphous particles (Seki & Yamamoto 1980), or fractal aggregates (Beckwith & Sargent 1991).

Since the dust is expected to be cold, the Rayleigh-Jeans limit is applicable to the millimeter emission. The flux in the optically thin and Rayleigh-Jeans regime, $F_\nu$, is given by

$$F_\nu = (2kT/D^2)(\nu/c)^2 \kappa M \tag{2}$$

where k is Boltzmann's constant, T is the temperature, D is the distance, c is the speed of light, and M is the mass. Combining measurements at two frequencies, it is possible to estimate the power law index from the flux at different frequencies

$$\beta = -2 + \log(F_{v1} / F_{v2}) / \log(v_1 / v_2) \tag{3}$$

As a function of both $F_{v1}$ and $F_{v2}$, the uncertainty in β depends on the uncertainty in both flux measurements and is given by

$$\delta\beta = 1/[(\ln(10)\log(v_1/v_2)] \{ (\delta F_{v1}/ F_{v1})^2 + (\delta F_{v2}/ F_{v2})^2 \}^{1/2} \tag{4}$$

based on the discussion in Taylor (1997).

    The OVRO data measure the flux at 3mm (112 GHz) and previous single dish measurements (André & Montmerle 1994) have been made at 1.3mm (240 GHz). Both values are used to estimate the dust opacity index for each primary given in Table 4. Possible systematic effects on the β value include contamination from the cloud in single dish data, which would make the 1.3mm flux too high, or flux that is resolved out by the interferometer which would make the 3mm flux too low; both of these effects artificially increase β, making the dust opacities given in Table 4 upper limits. For example, one of the targets – SR 24 – was observed with both a single dish (André & Montmerle 1994) and interferometer (Andrews & Williams 2005) at 1.3mm and the flux difference lowers the β for that source from 0.85 to –1.0. Because none of the secondaries were detected at 3mm, the unresolved 1.3mm flux is assigned entirely to the primary star. Another possible contribution to the millimeter flux is free-free emission that dominates the centimeter emission. Of our targets, only IRS 43 has published data at 3.6 cm and 6 cm (e.g. Girart et al. 2004); from the VLA data, the free-free component was estimated and subtracted from the millimeter flux. This correction is listed in Table 4, and the large uncertainty in the free-free contribution results in a large β uncertainty for IRS 43. If any free-free emission is present in the other sources it will be a larger fractional contribution of the measured flux at 3mm than at 1.3mm, given the spectral index of free-free emission, making the true value of β lower than calculated here. The uncertainty in β includes both rms noise and the absolute flux level uncertainty. By assuming a normalization for equation (1) of 0.1 cm$^2$/g at a frequency of 1000 GHz (Hildebrand 1983) and the β calculated for each target, the dust opacity at the observation frequency of 112 GHz is determined for each circumprimary disk.

    Three of the four Ophiuchus disks exhibit a dust opacity scaling with frequency of less than 2.0. The derived dust opacity power law index and mass for each circumprimary disk in the Ophiuchus systems is similar to previous measurements of young stars. For the Ophiuchus circumprimary disks, β ranges from 0.11 to 2.2. These results are similar to previous values determined from multi-wavelength photometry of Taurus and Orion targets over the range 0.6mm – 1.1mm (Beckwith & Sargent 1991), 0.35mm – 2mm (Mannings & Emerson 1994), and 0.8mm – 2.7mm (Dutrey *et al.* 1996). Most previously reported values of β are concentrated near unity, and these Ophiuchus binaries are also near the sharp peak of the distribution. Although there is not a unique explanation for the shallower power law index measured for the Ophiuchus targets, several possibilities include grain growth within the disks or differences in the composition or structure of the grains. If the low β is due to grain growth, then the similarity of the Class I and Class II disks suggests that this process has already occurred by the less-evolved Class I stage.

## 5.2 Disk Masses

Lower limits on the disk masses can be obtained from the 3mm flux and dust opacity with the assumption that the material is optically thin with the relation:

$$M_D[M_{sun}] \geq (F_{3mm}[mJy]\ D^2[pc])\ /\ (6.4 \times 10^7\ T[K]\ \kappa[cm^2/g]). \tag{5}$$

The distance is taken to be 160 pc as explained in Section 2. Two values of temperature, 15K and 30K, are used to calculate the disk mass from the OVRO primary disk flux or 3σ secondary disk upper limit and the calculated value of β. Because the dust opacity power law index for the secondary cannot be determined, the β calculated for the primary is used to estimate the secondary limit. The resulting circumprimary disk masses and circumsecondary disk mass limits are listed in Table 5; the uncertainty incorporates both the flux and β uncertainties and is given by the relation

$$\delta M_D = M_D\ \{\ (\delta F_{3mm}\ /\ F_{3mm})^2 + [-\delta\beta\ \ln(\nu[GHz]\ /\ 10^3)\ ]^2\ \}^{1/2}. \tag{6}$$

Upper limits on the ratio of the disk masses (calculated from the ratio of the secondary flux limits to the primary flux measurements) are also listed in Table 5 since numerical models of binary formation make predictions about the disk mass ratio. Two cases are considered for the secondary limits: the same temperature for the primary and secondary disk, and secondary disks 20% cooler than primary disks. A secondary stellar effective temperature 20% lower than the primary is the average of the two Class II sources; the Class I secondary effective temperatures are unknown. Table 5 reports the disk masses in units of the Minimum Mass Solar Nebula, the 0.01 Msun amount of material with solar abundance required to form the planets in the Solar System (Weidenschilling 1977). Differences in temperature cannot account for the entire difference in primary and secondary disk masses. In some cases the circumprimary disk masses are comparable to or greater than the Minimum Mass Solar Nebula, while the circumsecondary disk masses (listed as a fraction of the primary) are significantly lower.

A histogram of disk masses determined from large single-dish surveys of Taurus (Beckwith et al. 1990) is given in Figure 7 and the estimates of the Ophiuchus circumprimary disk masses are also indicated above the histogram as points or double-headed arrows. The submillimeter properties of Taurus and Ophiuchus have been found to be very similar (Andrews & Williams 2007). Among the Taurus sample, binary systems are differentiated, although the single dish measurements would not have resolved the individual components. The secondary disk mass upper limits are not plotted, nor are the upper limits from the single dish measurements. The range of masses plotted for each Ophiuchus binary corresponds to the temperature range of 15 K to 30 K. For SR 24 the masses derived from combining the OVRO flux with the IRAM 30m single dish or the SMA interferometer flux are both plotted and are substantially different. The Ophiuchus disk masses overlap the majority of the Taurus distribution, suggesting that the presence of a wide companion does not reduce substantially the primary disk mass relative to single stars with detected disks.

## 5.3 Evolution of Circumstellar Material in Binaries

When combined with previous maps of younger binaries in the same Ophiuchus star-forming region, the OVRO 3mm continuum maps complete an evolutionary sequence for widely-separated binaries from Class 0 through Class I to Class II. The cumulative data provide an observational constraint on the lifetime of circumstellar material in young binaries. All the systems discussed have separations in the hundreds of AU. This separation range is based on the limits of the mm arrays and also corresponds to physical separations exceeding the size of individual disks (e.g. Mannings & Sargent 2000)

The youngest Class 0 Ophiuchus binaries VLA 1623 and IRAS 16293-2422 reveal mm emission from both circumprimary and circumsecondary disks in addition to circumbinary material (Wootten 1989, Looney *et al*. 2000). The total masses of the circumstellar material around the primary/secondary components are estimated to be 0.04Msun/0.04Msun (VLA 1623) and 0.49Msun/0.61Msun (IRAS 16293-2422) (Looney et al. 2000) which would have been easily detectable in our OVRO data. The amount of that total mass confined to a disk relative to the envelope is difficult to quantify, but a rough upper limit can be obtained by dividing the total mass by the ratio of flux observed on small vs. large spatial scales. Using spatial frequency values of 5 and 50 k$\lambda$ to represent these scales results in circumprimary/circumsecondary disk masses of 0.02Msun/0.02Msun and 0.07Msun/0.2Msun, still detectable in the OVRO data. Since Class 0 systems experience significant infall of envelope material (e.g. Velusamy & Langer 1998, Velusamy et al. 1995) and reveal two mm detections (e.g. Looney et al. 2000, Launhardt 2004), the Class 0 maps permit the possibility of disks developing around each star. By contrast, the maps presented here of the Class I Ophiuchus binaries L1689SNO2 and IRS 43 with similar separations only detect mm emission from the circumprimary disk. Differences in sensitivity cannot explain the lack of two detections since the rms noise level in the OVRO Class I/II maps is slightly lower than in the corresponding BIMA Class 0 maps. This result suggests that the disk encircling the secondary is substantially depleted in mass by the Class I stage. The circumprimary disks detectable at the Class I stage persist in the Class II systems SR 24 and Elias 30, and these disks retain masses near the Minimum Mass Solar Nebula. SR 24 was previously observed at 1.3mm with the IRAM 30m, yielding results similar to ours but with lower angular resolution (Nürnberger *et al*. 1998), and with the SMA (Andrews & Williams 2005).

In addition to the Class 0 binaries observed in Ophiuchus, mm interferometry maps have been made of Class 0 binaries in other star-forming regions and in Bok globules, and typically emission is detected from both components (Launhardt 2004, Looney *et al*. 2000). From the observations of different evolutionary states, it is also possible to compare the disk mass ratios and mass ratio limits of the different classes. Launhardt et al (2004) examined Class 0 disk mass ratios (*q=secondary disk mass/primary disk mass*) and found that that 50% of the systems have *q>0.3* and 80% have *q>0.2*. These Class 0 data were taken with the same OVRO array and possess similar sensitivity as the observations presented here. Since the disk mass ratio limits for the Ophiuchus binaries, listed in Table 5, are comparable to the Class 0 detections, the differences seen in the maps of the Class I and II systems relative to the Class 0 systems is most likely a consequence of *lower secondary disk masses* rather than simply a decrease in sensitivity. Lower temperatures for the secondary disks are also unlikely to explain the 3mm nondetections of the Class I companions as discussed in the previous section.

The current results of the Ophiuchus Class I and Class II binaries are similar to those obtained in a previous examination of wide Class II binaries in the Taurus star-forming region (Jensen & Akeson 2003). In the Taurus study – which also had sufficient resolution to separate the two components – the distribution of millimeter dust emission in three of four pairs was restricted to the primary star. The disk mass ratio detection limits were also similar to the values for the Ophiuchus systems, listed in Table 5. Although the current sample is small, combining the Ophiuchus data with the results of Class II Taurus studies (Jensen & Akeson 2003, Jensen *et al*. 1996) constructs a nearly complete sample of mm-detected binaries with separations of ~3"-10" (only HN and HP Tau have not been observed). All of these seven Class II binaries exhibit dominant emission from the primary, with only two secondaries detected, both at a weaker level. This same trend is evident in the initial two Class I systems observed. The observed systems represent the subset of the binary population with the widest separations; the separation distribution peaks at 0".6 (Patience *et al*. 2002a) corresponding to tens of AU, but spatially resolved observations of subarcsecond systems will require CARMA and/or ALMA observations which will have longer baselines, greater sensitivity, and operate at shorter wavelengths.

**5.4 Disks and Extrasolar Planets in Binary Systems**

Since disks around young stars contain the raw materials for future planet formation, the OVRO results can be compared with the extrasolar planet population. In particular, our results suggest that planets might be expected to be present around primaries with wide companions and may be rare around secondaries. Initial stellar companion searches to known extrasolar planet stars revealed that several are primary stars with companions as close as ~50 AU (Luhman & Jayawardhana 2002, Patience *et al*. 2002b), indicating that planet formation can proceed in a circumprimary disk which is consistent with our results of massive, long-lived disks around younger primaries. More extensive surveys have identified further examples of extrasolar planet host stars that are primaries (e.g. Raghavan *et al*. 2006, Mugrauer *et al*. 2007).

Except for specialized techniques (e.g. Konacki 2005), radial velocity searches avoid binaries close enough for light from both objects to enter the spectrograph slit; however, some wider systems have been included in planet search programs. To assess the frequency of planets around primaries, the complete sample, including non-detections, needs to be known, so we cross-referenced a published 889-star subsample of an extensive radial velocity planet search program (Nidever *et al*. 2002) with the CCDM catalogue (Catalog of Components of Double and Multiple Stars; Dommanget & Nys 2002) to estimate the number of primaries with separations comparable to the OVRO systems that were searched for planets. Among the 889 sample stars, 111 matched the coordinates of CCDM primary stars, and only 25 matched the coordinates of secondaries, too few for an analysis of secondaries. Of the 111 primary stars, 4 have radial velocity planets. Although the numbers are currently small, the percentage of primary stars with radial velocity detected planets is 4% ± 2%, similar to the overall percentage of solar-type stars with planets (e.g. Marcy et al. 2000). This is consistent with our disk observations in suggesting that the presence of a wide companion does not seriously impact the ability to form planets.

**5.5 Comparison with Binary Star Formation Models**

The OVRO Ophiuchus observations also pertain to binary star formation theories. A variety of formation mechanisms have been proposed for binary stars including capture in small clusters with (McDonald & Clarke 1995) and without the effects of disks (Sterzik & Durisen 1998, McDonald & Clarke 1993), scale-free fragmentation (Clarke 1996), and accretion following fragmentation in a cluster (Bate 2001) or loose environment (Bate 2000). The measurements of relative disk masses can be compared with predictions from theoretical models. Figure 8 plots the disk mass ratio as a function of the stellar mass ratio for Class II binaries observed in this study and Taurus (Jensen & Akeson 2003); only Class II binaries are plotted, since their stellar mass ratios are well-determined. First, in the scale-free fragmentation scenario (Clarke 1996, Murray & Clarke 1993), the more massive primary should have the more massive disk, and this trend is seen in the Ophiuchus binaries. This agreement with the broad predictions of the model was also noted for Taurus Class II systems (Jensen & Akeson 2003).

The second comparison involves simulations that model the accretion of material onto protobinary disks (Bate 2000, Ochi *et al*. 2005). These models also predict the larger circumprimary disks that are observed, but further quantify the relationship between the mass ratio of the stars and the mass accretion rates of the disks. The relative accretion rate does not differentiate between material deposited onto the disk or star, but – if this is similar for primary and secondary – then the accretion ratio generated in the numerical studies provides an indication of the expected disk mass ratio measured with the observations; for the following comparison we use the accretion ratio as a proxy for the disk mass ratio. A series of calculations have modeled the relative accretion rate onto the secondary and primary with different conditions for the mass ratio of the stellar cores, core density profile, and core rotation (Bate 2000) and for different specific angular momentum of the infalling gas, stellar mass ratio, and numerical resolution (Ochi *et al*. 2005).

In one of the accretion models (Bate 2000), the minimum accretion ratios occur for accretion onto cores with uniform density profiles and with differential rotation inversely proportional to radius, while the maximum ratios result from simulations with cores in solid body rotation with density profiles that decline as $r^{-2}$. The SR 24 disk mass ratio limit is consistent with the differentially-rotating, uniform density core simulation, but inconsistent with the models of $r^{-2}$ density profile cores in solid body rotation. All other binaries have disk mass ratio limits lower than the model predictions, since the disk mass ratios are substantially less than the stellar mass ratios, as shown in Figure 8. The second suite of numerical simulations (Ochi *et al*. 2005) calculates secondary-to-primary mass accretion ratios of ~0.5-1.0 for $q_{star}$=(secondary star mass/primary star mass)=0.7-0.9 binaries and accretion ratios of ~0.3-1.0 for $q_{star}$=0.4-0.6 systems. The high stellar mass ratio Taurus binary with two disks is consistent with one of the models, and the remaining lower q pairs in Taurus and Ophiuchus have disk mass ratio limits lower than the predicted mass accretion ratio values.

Another model of disk evolution involving geometrically thin non-self-gravitating disks subject to viscous and magnetic influences (Armitage *et al*. 1999) can be compared with the Ophiuchus and Taurus results. This model follows the later stages after the formation of the binary and disk, more comparable to the Class II stage. The evolution of the disk mass ratio for a 50 AU binary is calculated with different initial primary and secondary mass accretion rates (Armitage *et al*. 1999). Since measurements of mass accretion rates derived from spatially resolved spectra of T Tauri binaries (White & Ghez 2001, Hartigan & Kenyon 2003) typically find higher mass accretion rates from the primary, we use that category of the simulations. The simulations calculate disk mass ratio evolutionary tracks for stellar mass ratios of 0.1, 0.19, and 0.44, and the tracks are plotted on Figure 8. The absolute time scaling depends on the value of the viscosity $\alpha$ parameter, chosen to be 0.005 for the models. Two of the Class II systems have stellar mass ratios within the considered range. The binary with the lower stellar mass ratio has a disk mass ratio limit consistent with ages greater than 1Myr, and the binary with the higher stellar mass ratio has a disk mass ratio limit that corresponds to an age of 2Myr – consistent with the ages expected for a Class II. The similarity of ages implied from stellar evolutionary models and disk evolutionary models suggest that the viscosity used in the model is representative of the actual value in the disk. Although the combined data for spatially resolved millimeter disk measurements is a small sample, these observations are uniquely suited for comparisons with models of binary disk masses; larger-scale infrared surveys of binaries that measure excesses attributable to inner disks cannot reliably estimate the disk mass.

**5.6 Comparison with Planet Formation Timescales**

The presence of relatively massive circumprimary disks, coupled with the very limited material in secondary circumstellar disks, can be compared with the expectations of planet formation models. The most interesting comparison of disk lifetimes and planet formation timescales requires an assessment of the ages of the Class I and II stages. Although the ages of young stars are subject to a number of uncertainties, Class II objects are commonly dated by comparing their luminosities and effective temperatures with theoretical evolutionary tracks (e.g. Palla & Stahler 1999, Baraffe *et al*. 1998, D'Antona & Mazzitelli 1997). Using this method and K-band spectroscopy of Ophiuchus Class II and III yields an age range of 0.1-1.0 Myr (Luhman & Rieke 1999). An analysis of Ophiuchus Class II members based on ISOCAM suggests an age of 0.5 Myr or up to 1-2 Myr (Bontemps *et al.* 2001). Similarly, large-scale photometric (Kenyon & Hartmann 1995) and spectroscopic (*e.g.* White & Ghez 2001) studies of Taurus determine ages of 1-2 Myr for the Class II objects.

Because the Class I sources are more embedded, it is much more difficult to detect photospheric features and to place the objects on evolutionary tracks; consequently, the Class I stage lifetime is typically inferred by assuming a constant star formation rate and measuring the proportion of Class I objects. With near-infrared data, corrected for reddening and complete to K<12mag, a population of Ophiuchus Class I objects only 17% of the Class II/III population was found, implying a Class I lifetime of ~0.075 – 0.15 Myr (Rieke & Luhman 1999). Similar results were obtained from a statistical study of Taurus in which the Class I population represented only 13% of the combined Class II/III numbers, suggesting Class I objects have ages of 0.1-0.2 Myr (Kenyon & Hartmann 1995). In summary, Class I ages based on population studies are significantly less than 1Myr which is an important timescale for planet formation models, as described below. Further support for the young age of our Class I objects comes from infrared spectra of the primary of IRS 43 which reveal photospheric lines and allow estimates of the stellar luminosity and temperature to be compared with evolutionary models. Analysis of the infrared spectrum of IRS 43 suggests an age of 0.1-0.2 Myr (Greene & Lada 2002, Prato *et al.* 2003) for this Class I object, consistent with the young population-based estimates of the Class I stage lifetime. Similar spectra do not exist for L1689 SNO2, so we rely on the statistical age of Class I objects for that source.

Disk lifetimes are an important constraint on planet formation models, as the disk must survive long enough for planets to form. Combining our results with a previous millimeter interferometry study of Class II binaries (Jensen & Akeson 2003) provides significant evidence that massive outer secondary disks dissipate by the Class II stage (only 1 of 6 secondary disks detected), and there is initial evidence from our Class I data suggesting that this decline in disk mass has already occurred by the earlier Class I stage. Taking an age of ~0.2Myr as representative of the Class I stage based on Ophiuchus population studies and the IRS 43 primary results, the limited lifetimes of massive circumsecondary disks suggested by our observations may negatively impact the potential to form planets around secondary stars in binary systems. One class of giant planet formation models involves a gradual accumulation of icy planetesimals over a period of ~1Myr to form a solid core and a subsequent rapid capture of a gaseous envelope (*e.g.* Bodenheimer & Pollack 1986). The disk lifetimes for circumprimary disks implied by the observations are consistent with this scenario, but the <1Myr lifetime for circumsecondary disks may be insufficient for planet formation by this mechanism. Alternatively, a gravitational instability in the disk that occurs on a timescale of ~1000yr and results in a more rapid condensation to form a giant planet in ~0.1Myr (Cameron 1978, Boss 1997, 1998) could have occurred by the Class I stage. The significantly lower circumsecondary disk masses could result from either depletion caused by the formation of giant planets through a disk instability or a more rapid disk dissipation timescale.

If the ages of the Class I objects are instead better represented by the 1-2 Myr measured for Taurus Class I systems (White & Hillenbrand 2004), then there may have been significantly more time available for the disks to develop planetary systems. Although the timescale to reduce the disk mass significantly is still less than this age, a limit of 1-2 Myr is a less restrictive dissipation timescale than ~0.2 Myr and does not completely preclude the gradual accumulation of giant planet cores in time to capture a gaseous envelope. If the Class I and Class II stages are actually coeval, then the similarity in the distribution of dust emission and dust opacity scaling for the primary stars is consistent with this scenario.

## 6. Summary

With 3mm interferometric observations of two Class I and two Class II pre-main sequence binaries in Ophiuchus, the distribution of the dust emission between components, disk masses or limits, and primary dust opacity index have been determined. All systems show similar results, despite their different evolutionary stages, with the primary stars alone displaying detectable mm emission. This is in marked contrast to results from younger (Class 0) Ophiuchus binaries where emission is measured from both primary and secondary (Wootten 1989, Looney *et al*. 2000), but similar to observations of Class II Taurus binaries (Jensen & Akeson 2003). The combined results suggest a rapid dissipation of the cool, outer disk material around secondary stars such that the total disk mass is significantly reduced by the less than 1 Myr age estimated for Ophiuchus Class I systems, which may negatively impact planet formation around secondary stars. The more massive primary stars have the more massive disks, in broad agreement with binary formation models (Clarke 1996, Bate 2000, Ochi *et al*. 2005), though the upper limit on the disk mass ratio may be lower than expected from accretion simulations. Interestingly, the optical spectrum of one of the secondaries reveals signs of accretion of disk material onto the stellar surface, indicating that at least some secondaries still retain inner disks with very limited mass.

The circumprimary disk masses are comparable to previous determinations of disk masses for both single and binary stars in the Taurus region (Beckwith *et al*. 1990). Additionally, the circumprimary disk masses are comparable to or exceed the 0.01Msun value of the Minimum Mass Solar Nebula (Weidenschilling 1977) in most cases, suggesting that the binary environment – with companions separated by 100's-1000 AU – may not be hostile to planet formation for the primary star. Among stars with known radial velocity planets, some are known to have companion stars (e.g. Mugrauer *et al*. 2007). The dust opacity index for each primary disk is within the range of previous estimates of T Tauri star disks and smaller than expected for interstellar dust grains. One possible explanation is grain growth within the disks (Miyake & Nakagawa 1993) and the similarity between the Class I and Class II values implies that the coagulation and accumulation of dust grains has progressed significantly by the Class I stage.

**Acknowledgements**
The Owens Valley Millimeter Array is operated by the California Institute of Technology under funding from the National Science Foundation. EJ gratefully acknowledges support from NSF grant AST-0307830. Funding for J.P. was provided by a Michelson Fellowship from the Michelson Science Center and the NASA Navigator Program. This work made use of the SIMBAD database operated by CDS, France, the NASA Astrophysics Data System, and data products from the Two Micron All Sky Survey, which is a joint project of the University of Massachusetts and the Infrared Processing and Analysis Center/California Institute of Technology, funded by the National Aeronautics and Space Administration and the National Science Foundation. We thank members of the OVRO group for valuable discussions on data analysis: H. Arce, J. Carpenter, J. Koda, C. Sanchez-Contreras, and K. Sheth; and we thank H. Arce, M. Bate, L. Looney, R. White, and A. Sargent for productive discussions of the scientific results.

Tables

Table 1: Properties of the Observed Ophiuchus Systems

| Target[a] | Class[b] | R.A. (2000)[c] | Dec. (2000)[c] | Coordinate error ellipse (RA x Dec)[c] | Separation[d] | Angle[d] | Prim./Sec. Spectral Types[e] | Stellar Mass Ratio ($M_{sec}/M_{prim}$) | Alt. Name |
|---|---|---|---|---|---|---|---|---|---|
| IRS 43 | I | 16:27:26.94 | -24:40:50.8 | 0".09 x 0".09 | 6".99 | 322 | K5 / ? | >0.2[f] | YLW 15 |
| L1689SNO2 | I | 16:31:52.11 | -24:56:15.7 | 0".07 x 0".06 | 2".92 | 240 | ? / ? | ? | GWAYL 5 |
| SR 24 | II | 16:26:58.51 | -24:45:36.9 | 0".08 x 0".06 | 6".00 | 348 | K2 / K8 | 0.3 | HBC 262 |
| Elias 30 | II | 16:27:10.28 | -24:19:12.7 | 0".06 x 0".07 | 6".70 | 175 | G2.5 / M4 | 0.06 | SR 21 |

Notes: (a) target names used in Haisch *et al.* 2002 or Reipurth & Zinnecker 1993; corresponding SIMBAD names for each source are: ROXR1 43, GWAYL 5, EM* SR 24, and Elia 2-30, (b) classification adopted for this study, details in Section 2 (c) from 2MASS All-Sky Survey (Cutri et al. 2003), (d) from Haisch *et al.* 2002 or Reipurth & Zinnecker 1993, (e) if there is more than one measurement, the higher spectral resolution estimate from Prato *et al.* 2003 or Greene & Lada 2002, (f) assuming the primary mass estimated from higher resolution spectroscopy (Greene & Lada 2002) and the companion is stellar ($M_{sec}>0.08M_{sun}$)

Table 2: OVRO Position Measurements

| Target | Beam Size | Beam Angle | R.A. (2000) mm Peak | Dec. (2000) mm Peak | Uncertainty RA x Dec.[a] mm Position | Uncertainty RA x Dec.[b] mm Position | Coincident with primary? |
|---|---|---|---|---|---|---|---|
| IRS 43 | 5".1 x 3".1 | -3.6° | 16:27:26.93 | -24:40:50.0 | 0".1 x 0".06 | 0".4 x 0".2 | Y |
| L1689SNO2 | 4".5 x 2".8 | -2.3° | 16:31:52.11 | -24:56:16.2 | 0".09 x 0".06 | 0".4 x 0".3 | Y |
| SR 24 | 4".5 x 2".9 | -7.3° | 16:26:58.51 | -24:45:36.5 | 0".09 x 0".06 | 0".2 x 0".1 | Y |
| Elias 30 | 6".6 x 4".6 | 8.9° | 16:27:10.23 | -24:19:12.7 | 0".1 x 0".09 | 0".8 x 0".6 | Y |

Notes: (a) 2% of the beam, based on the small angular distance between source and calibrator (<1.5degrees) and the uncertainty in the baseline solution, ~0.2λ, (b) beam size divided by primary disk SNR in final map

Table 3: OVRO Flux and Size Measurements

| Target | Primary $F_{3mm}$ (mJy) | Map noise 1σ (mJy) | Flux Calibration uncertainty 1σ (mJy) | Secondary $F_{3mm}$ Limit (mJy) | Deconvolved Angular Size |
|---|---|---|---|---|---|
| IRS 43 | 15 | 1.7 | 1.7 | <5.2 | 2".7 x 2".0[a] |
| L1689SNO2 | 8.6 | 1.6 | 0.9 | <4.7 | Point |
| SR 24 | 32 | 1.8 | 3.5 | <5.5 | 3".1 x 2".4 |
| Elias 30 | 6.1 | 2.0 | 0.7 | <5.9 | Point |

Notes: (a) this source is only marginally resolved, since the size is only half the N-S beam size

Table 4: Dust Opacity Scaling

| Target | $F^{IRAM}_{1.3mm}$ (mJy)[a] | $F^{ff}_{1.3mm}$ (mJy)[c] | $F^{OVRO}_{3mm}$ (mJy)[d] | $F^{ff}_{3mm}$ (mJy)[c] | $\beta_{primary}$ |
|---|---|---|---|---|---|
| IRS 43 | 75 ± 8 | 14 ± 5 | 15 ± 2.4 | 10 ± 3 | 1.3 ± 1.0 |
| L1689SNO2 | 150 ± 8 | | 8.6 ± 1.8 | | 1.8 ± 0.3 |
| SR 24 | 280 ± 8 | | 32 ± 4.0 | | 0.9 ± 0.2 |
| | 68 ± 2[b] | | 32 ± 4.0 | | -1.0 ± 0.2 |
| Elias 30 | 150 ± 8 | | 6.1 ± 2.2 | | 2.2 ± 0.5 |

Notes: (a) all values from single dish measurements reported in André & Montmerle (1994) except as noted, (b) measurement from SMA (Andrews & Williams 2005), (c) free-free emission extrapolated from

VLA measurements at 3.6 cm and 6cm (Girart *et al*. 2004); the free-free contribution is subtracted before calculating $\beta$, (d) uncertainty includes both the rms map noise and absolute flux calibration uncertainty

Table 5: Disk Masses and Limits

| Target | T=15K Primary Disk Mass (0.01 Msun) | T=30K Primary Disk Mass (0.01 Msun) | $T_{sec}=T_{prim}$ Disk Mass Ratio Limit $(Disk_{Sec}/Disk_{Prim})^b$ | $T_{sec}=0.8T_{prim}$ Disk Mass Ratio Limit $(Disk_{Sec}/Disk_{Prim})$ |
|---|---|---|---|---|
| IRS 43 | 2.2 ± 5.2 | 1.1 ± 2.6 | <0.3 | <0.4 |
| L1689SNO2 | 11 ± 6.9 | 5.3 ± 3.5 | <0.5 | <0.6 |
| SR 24 | 5.4 ± 2.0 | 2.7 ± 1.3 | <0.2 | <0.3 |
|  | 0.093 ± 0.036[a] | 0.047 ± 0.018[a] | <0.2 | <0.3 |
| Elias 30 | 20 ± 20 | 10 ± 10 | <1.0 | <1.3 |

Notes: these mass estimates are given in units of the Minimum Mass Solar Nebula and assume the standard Ophiuchus distance of 160 pc; if the distance is 140 pc, then the values should be 23% lower, (a) calculated using 1.3mm flux from SMA rather than IRAM single dish flux (b) calculated from the ratio of the secondary flux limit and the primary flux (listed in Table 4)

Figure Captions

Figure 1a-d: The 24"x24" OVRO maps for each source – (a) IRS 43, (b) L1689SNO2, (c) SR 24, and (d) Elias 30 – are shown as a contour plot overlay on the 2MASS $K_s$ image. For the system without spectral types, the primary star is assumed to be the brightest K and L source. The solid contours plotted are in increments of the RMS noise level (listed in Table 3) starting at 2 times and extending to 4-12 times, depending on the source. The dashed contours are –2 and –3 times the RMS noise level. The alignment of the 2MASS and OVRO maps is based on the absolute positions.

Figure 2a-c: Maps of the CO spectral line data for IRS 43 are shown: (a) $^{13}CO(1-0)$, (b) $CO(1-0)$, and (c) $C^{18}O(1-0)$. The velocity ranges corresponding to the red and blue contours, respectively, are: (a) 5.3-8.6 km s$^{-1}$ and 0.1-2.1 km s$^{-1}$, (b) 5.8-6.4 km s$^{-1}$ and 1.0-2.5 km s$^{-1}$, and (c) 4.4-5.7 km s$^{-1}$ and 1.7-3.0 km s$^{-1}$. Each map is 40"x40" in size and solid contours denote blue-shifted emission, while dashed contours mark red-shifted emission. Velocity ranges for the red and blue-shifted lobes are different for each map and are given in section 4.2. The contour levels in Figure 2a are 3,5,7,9 times the average RMS noise level of the blue-shifted and red-shifted map, 6.0x10$^{-2}$ Jy beam$^{-1}$ km s$^{-1}$. For Figure 2b the contours are 4,8,12,16,20 times the average RMS noise level of 8.9x10$^{-2}$ Jy beam$^{-1}$ km s$^{-1}$, and for the weaker $C^{18}O$ detection in Figure 2c the contours are 3,4,5 times the average RMS noise level of 6.7x10$^{-2}$ Jy beam$^{-1}$ km s$^{-1}$.

Figure 3: Channel plots for the IRS 43 CO(1-0) data are shown; each panel is the sum of four channels and covers a velocity range of ~1km/s. The maps are 20"x20" in extent. The top four panels cover the velocities in the red lobe of Figure 2b and the bottom four panels cover the blue lobe; the middle plots are velocities near the system value. The velocity listed in the corner of each plot is the low end of the velocity range of the four channels. The first contour level is four times the RMS noise level value of 8.9x10$^{-2}$ Jy beam$^{-1}$ km s$^{-1}$, and the steps are also increments of four times the RMS noise level.

Figure 4: Comparison of near-infrared and millimeter images of IRS 43. The background image is from HST/NICMOS with the F160W ($\lambda$ = 1.6 µm) filter with the pipeline calibration, and the OVRO CO(1-0) map is aligned and overlaid. The OVRO beam is shown in the lower left corner and the orientation is such that North is rotated 147° to the left of the top.

Figure 5: A 24"x24" map of the CO(1-0) emission of SR 24 is shown. The line emission integrated over 9 channels ranging in velocity from 5.8-7.6 km/s; no comparable detection is seen in the blue-shifted velocity channels. The contours are 3,4,5 times the RMS noise level of 6.0x10$^{-2}$ Jy beam$^{-1}$ km s$^{-1}$ and indicate a weak detection centered on the location of the secondary, which is itself a close 0".197 pair.

Figure 6a-d: High-resolution optical spectra of Hα emission in some of the target binaries. All spectra are smoothed with an 11-pixel boxcar, and normalized to the continuum. Both components of SR 24 show strong Hα emission, indicating high accretion rates. The SR 24 spectra are strikingly similar, despite the very different mm

emission of primary and secondary. Elias 30A shows weak but broad Hα emission, with a central reversal that descends below the continuum; note the compressed y scale in this panel. The Hα emission in L1689SNO2 A is narrower than in the other sources, and the continuum level is very uncertain (see text).

Figure 7: The Ophiuchus primary star disk masses are compared with a histogram of disk masses determined for Taurus sources from IRAM single dish 1.3mm measurements (Beckwith *et al.* 1990). The x-axis indicates the disk mass and the y-axis applies to the histogram and plots the number of Taurus disks of each mass bin. The OVRO-determined Ophiuchus disk masses are shown above the histogram for clarity and the range of masses associated with temperatures from 15K to 30K is plotted as a double-headed arrow. All the Taurus systems included in the plot have been searched for companions, and disk masses of binary systems are plotted with darker shading, though the single-dish measurements would not have resolved the individual stars.

Figure 8: The Ophiuchus and Taurus disk mass ratio upper limits and detection are plotted as a function of stellar mass ratio. The dashed lines are simulations (Armitage et al. 1999) of the evolution of binary disk mass ratios over time for several stellar mass ratios.

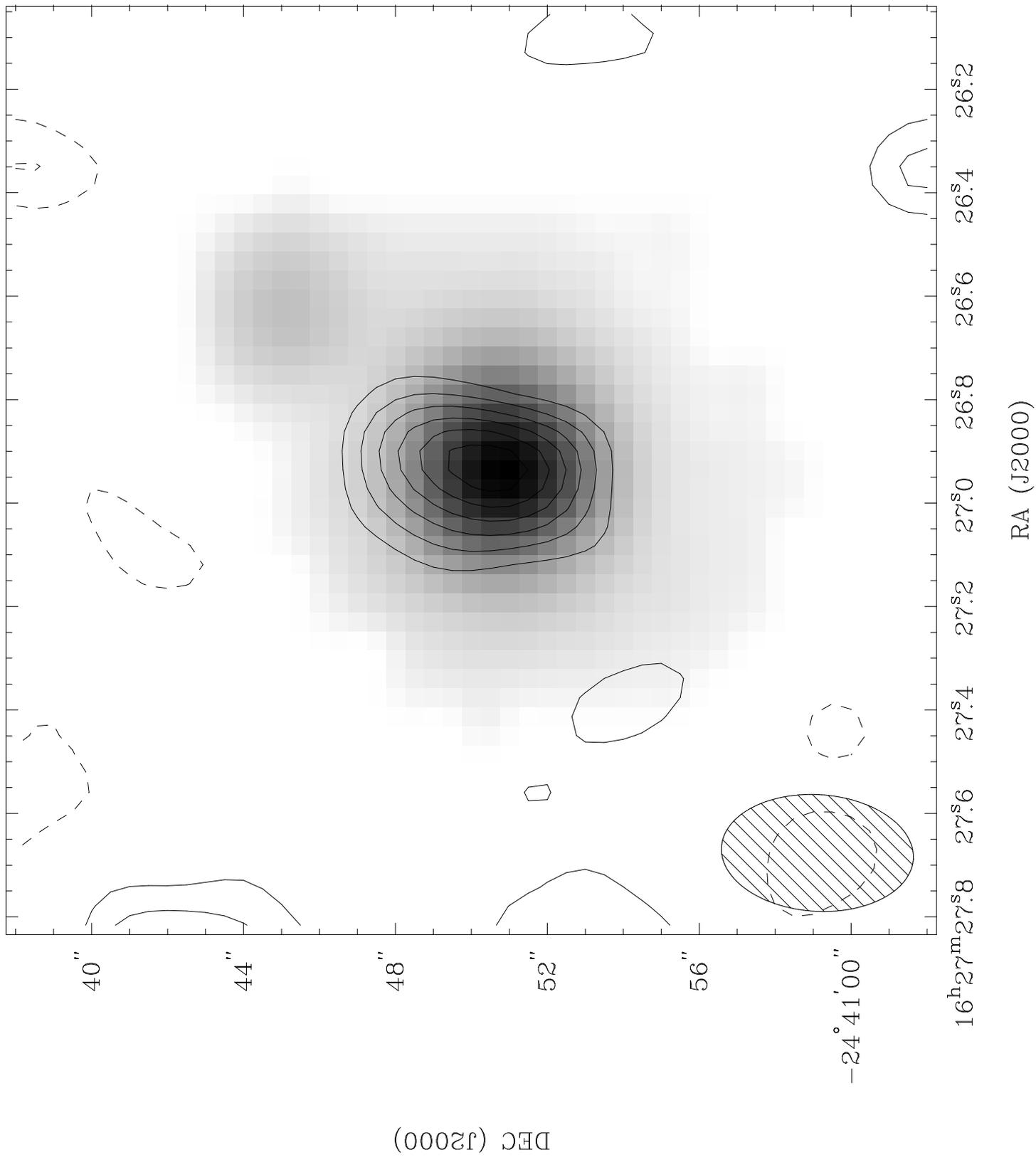

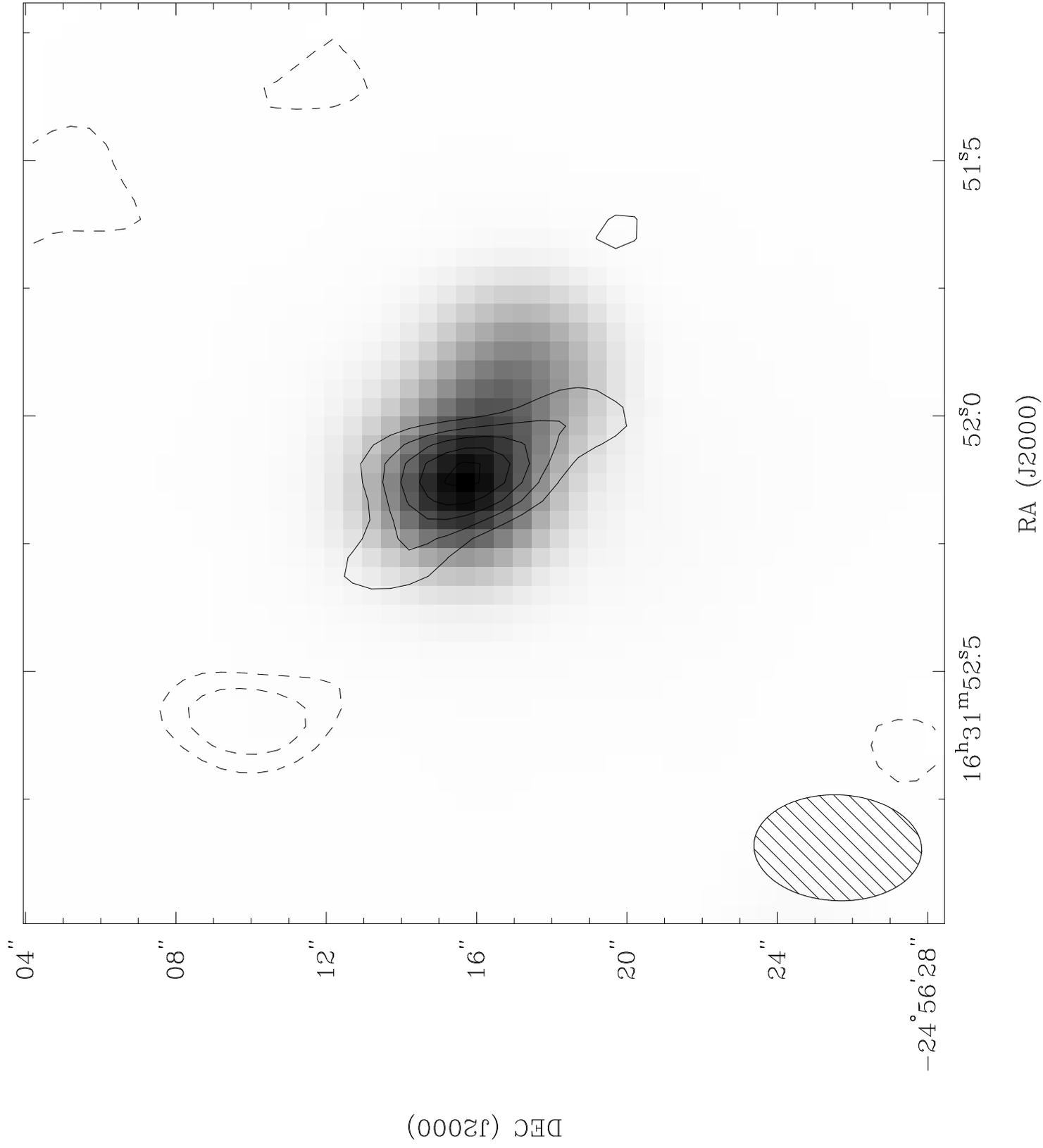

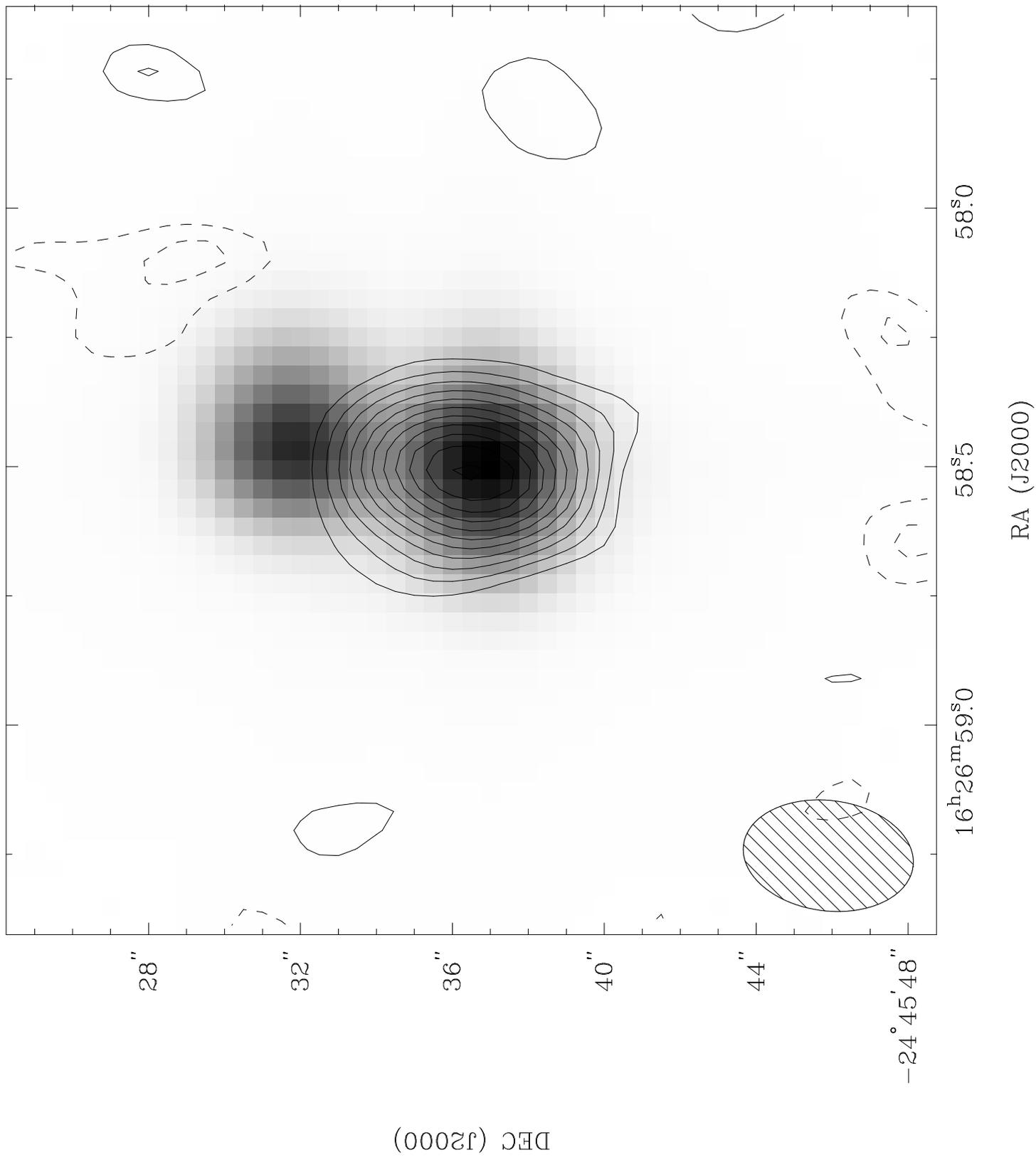

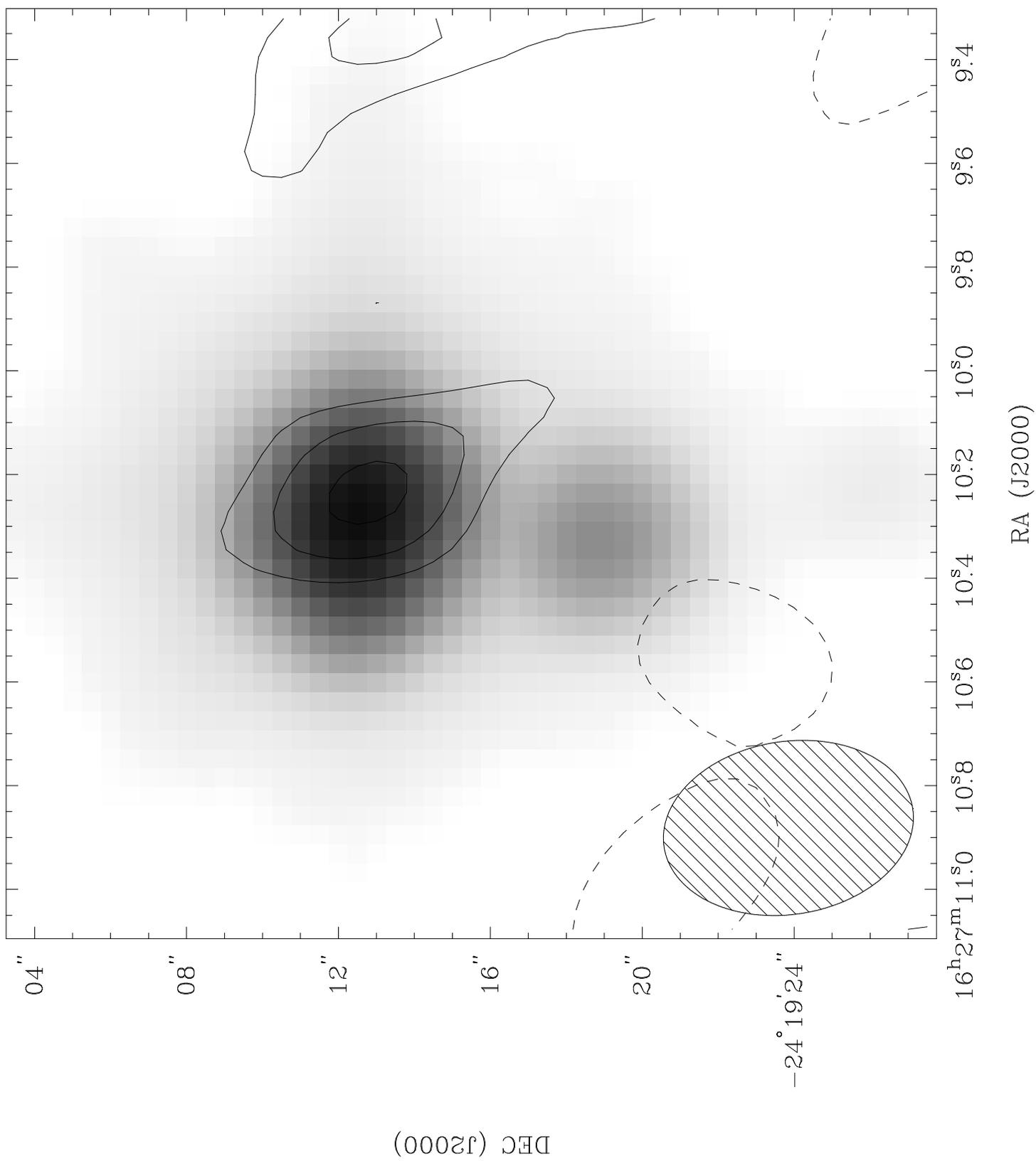

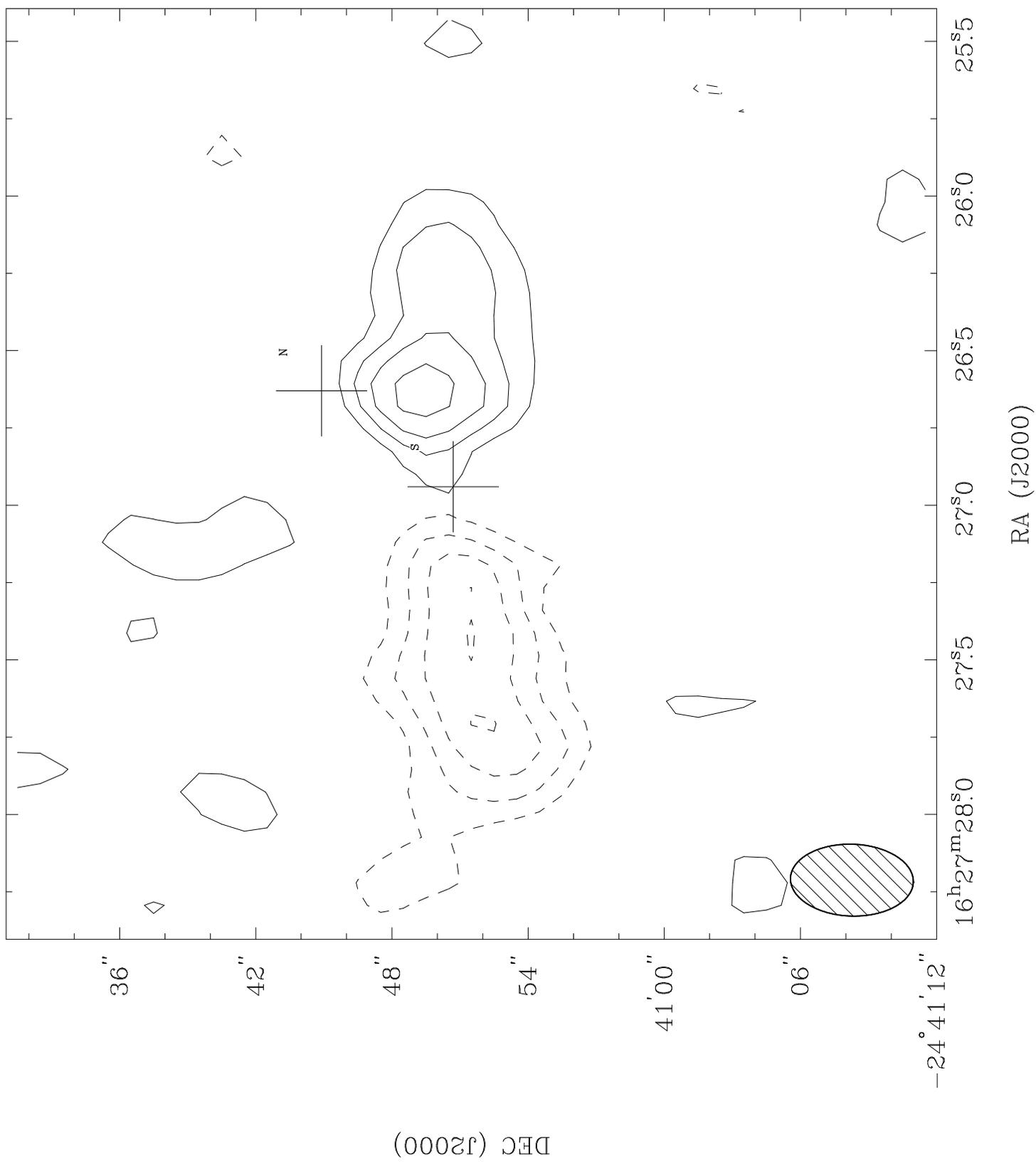

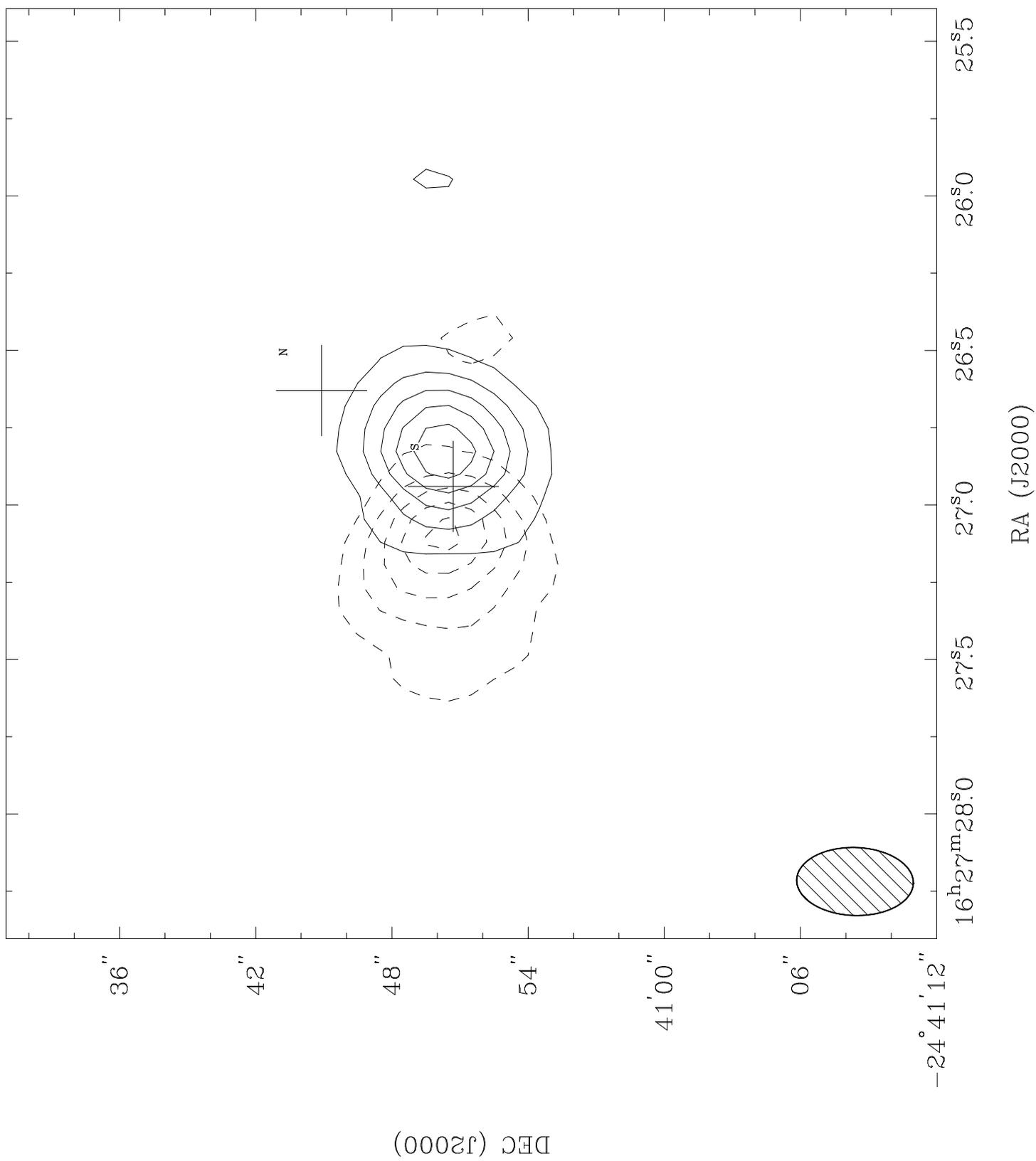

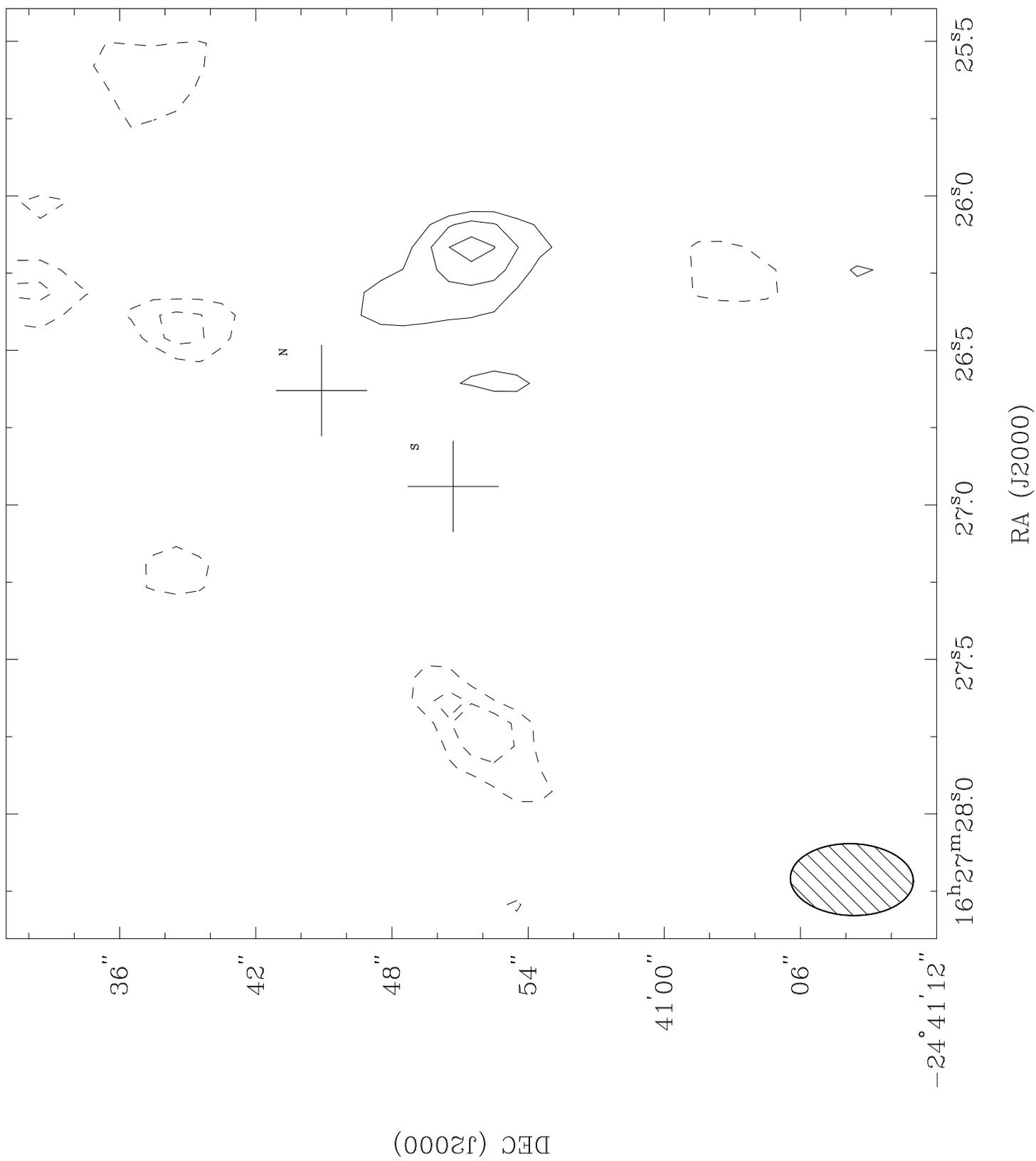

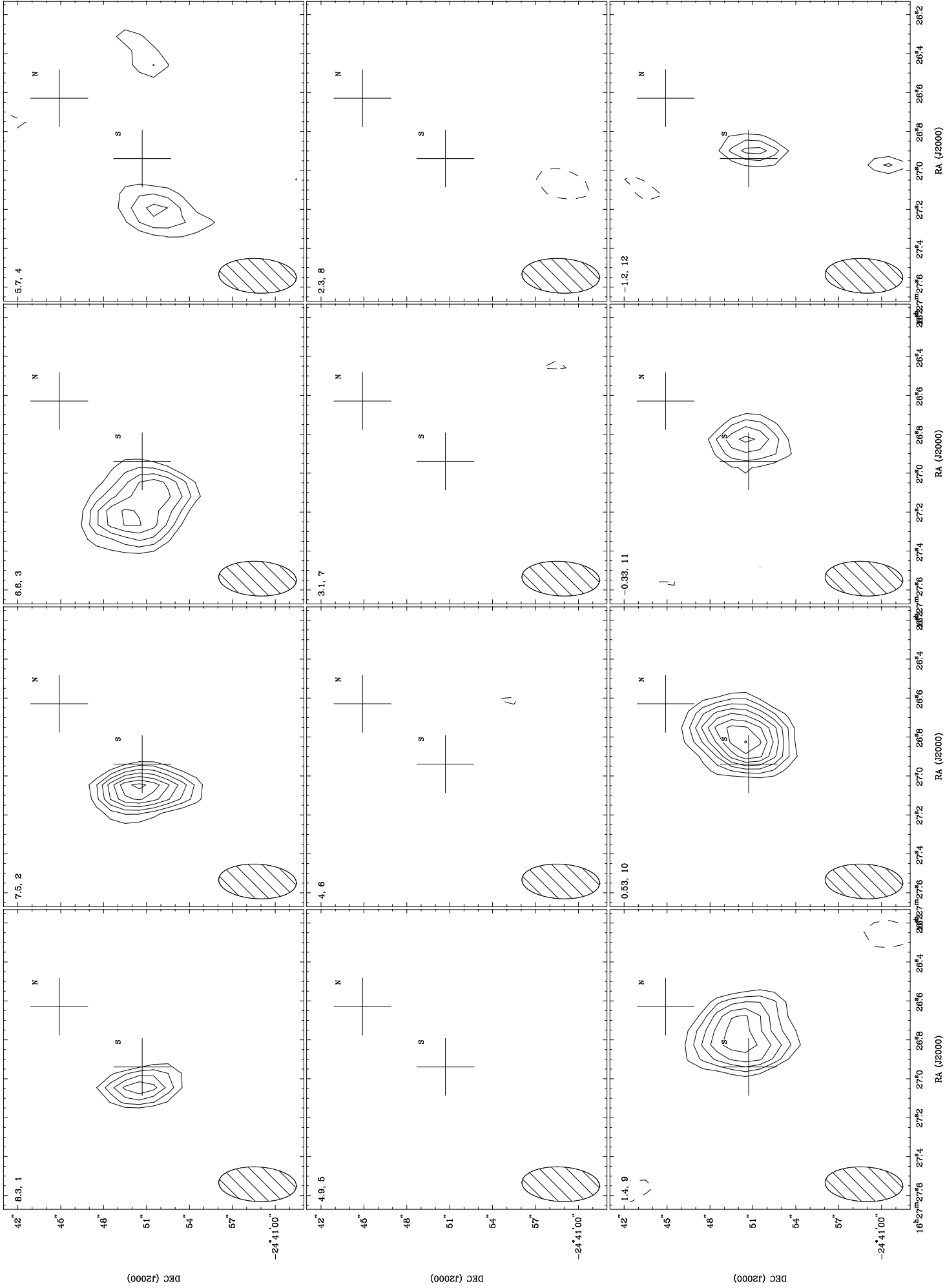

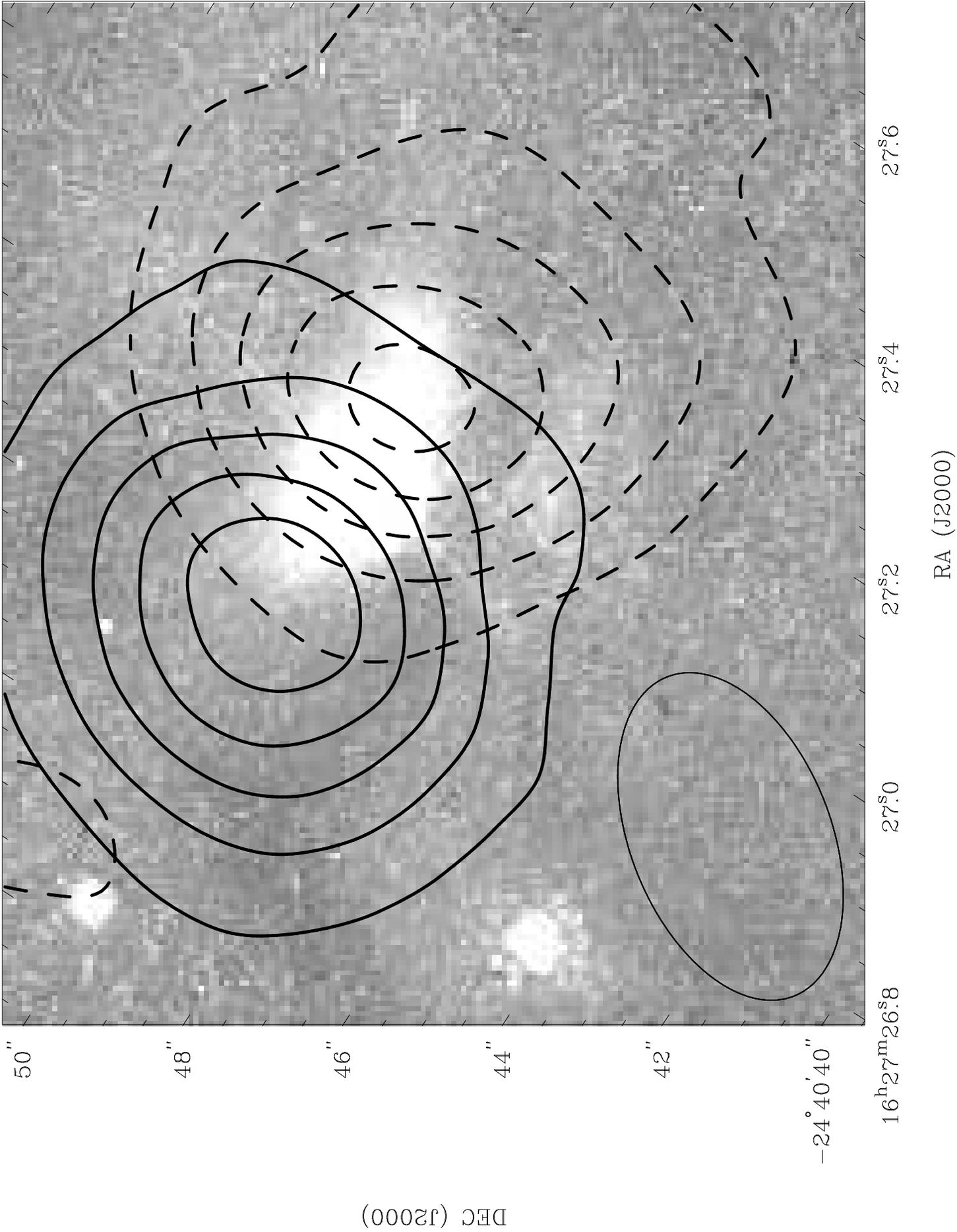

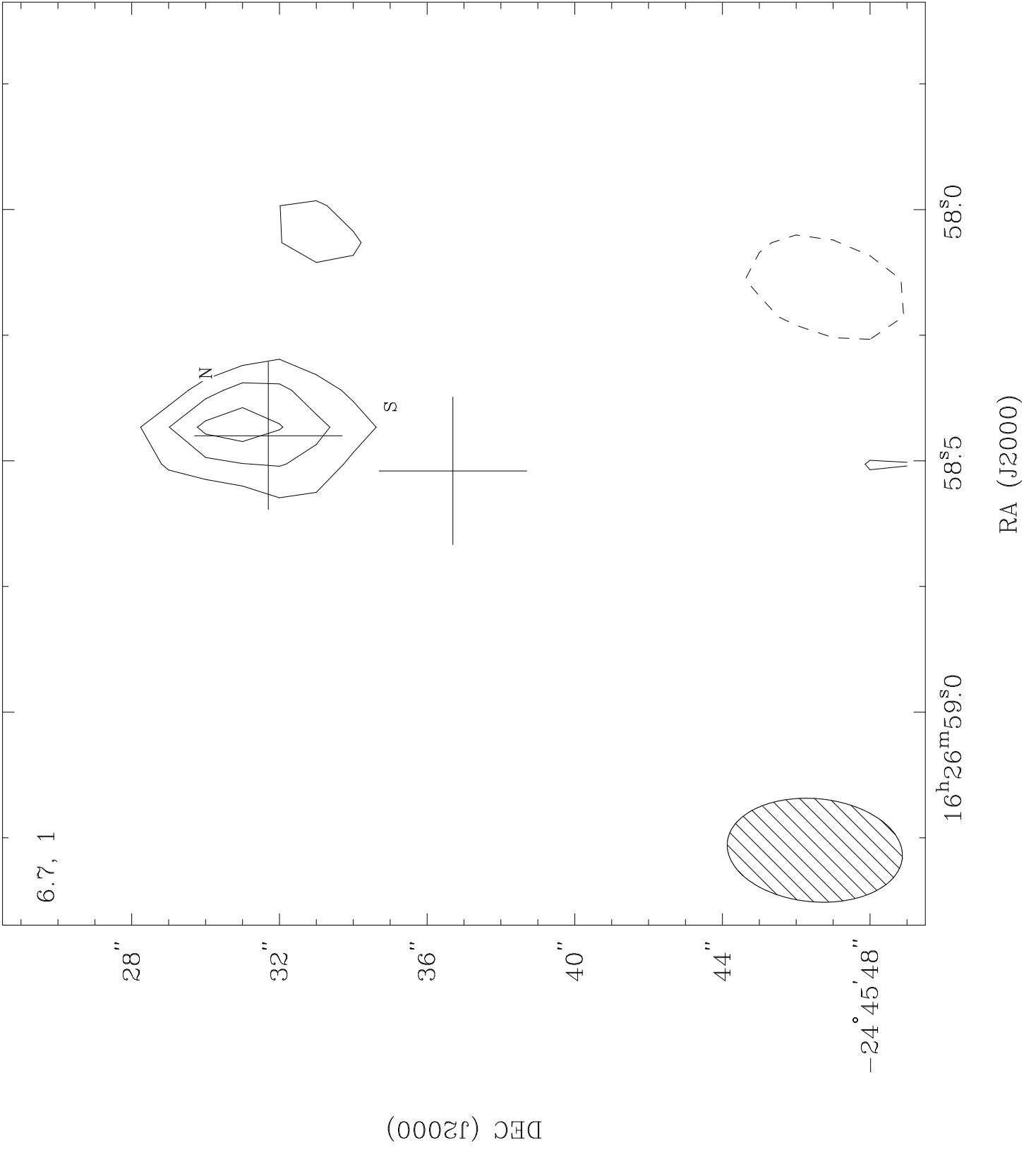

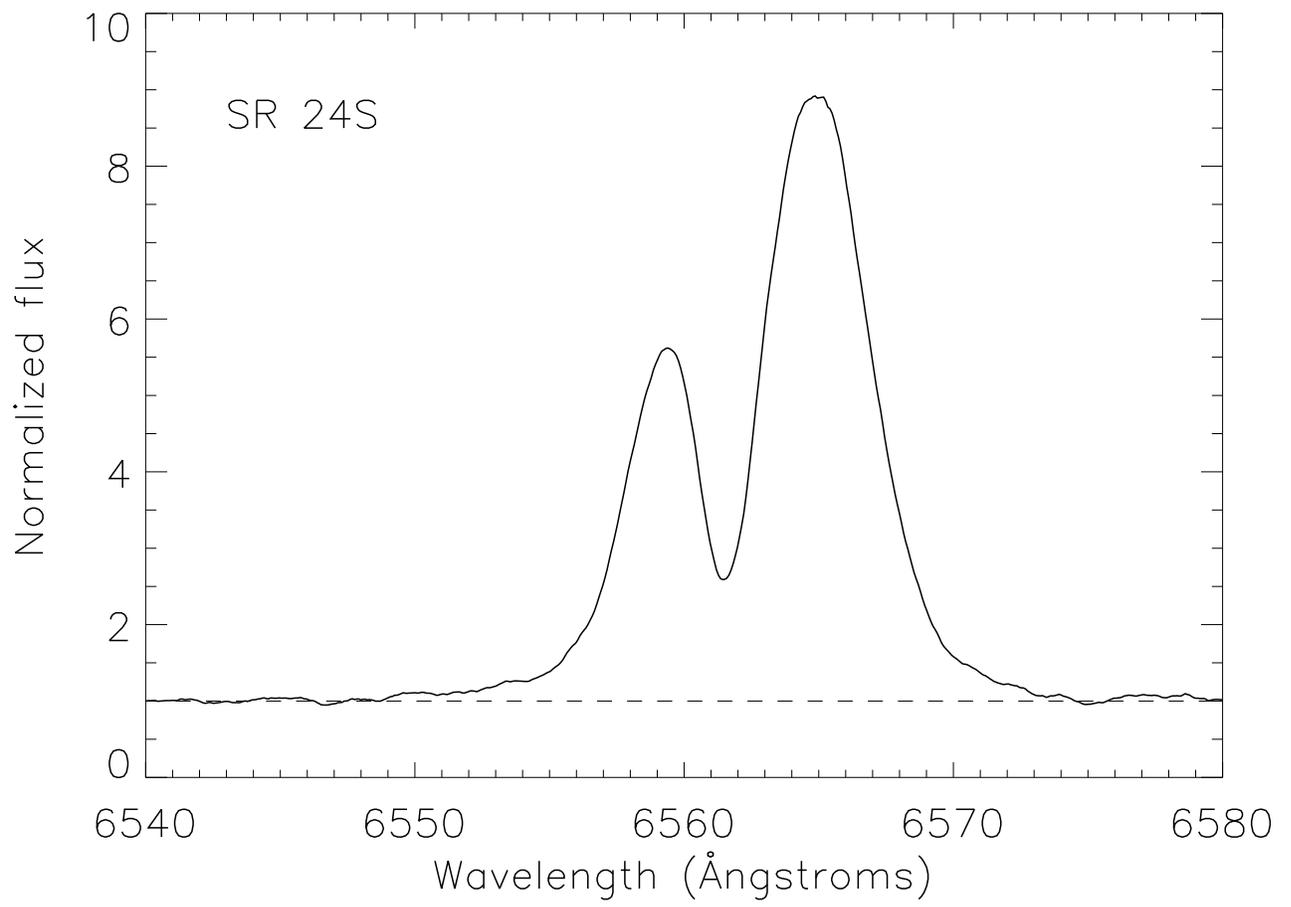

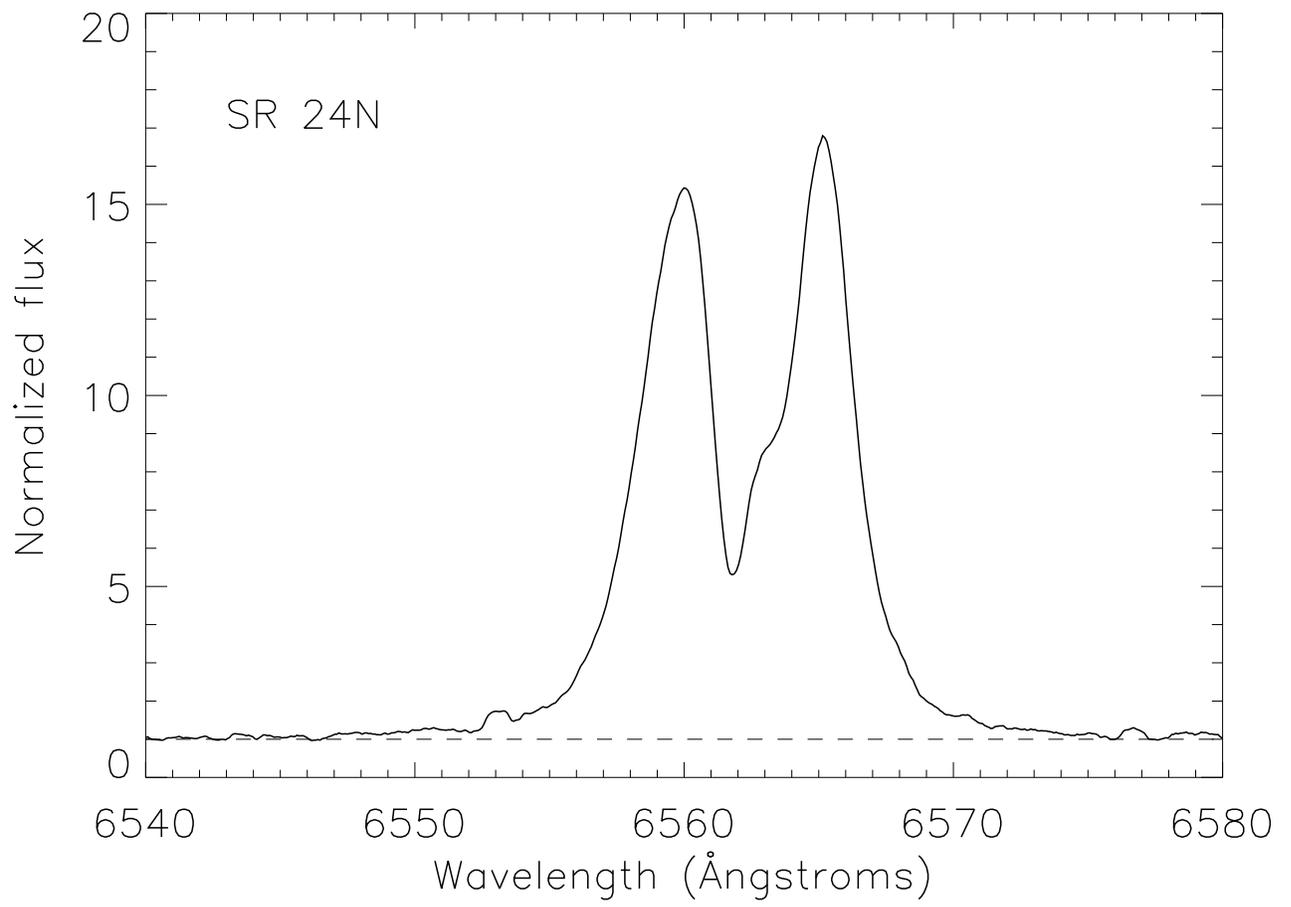

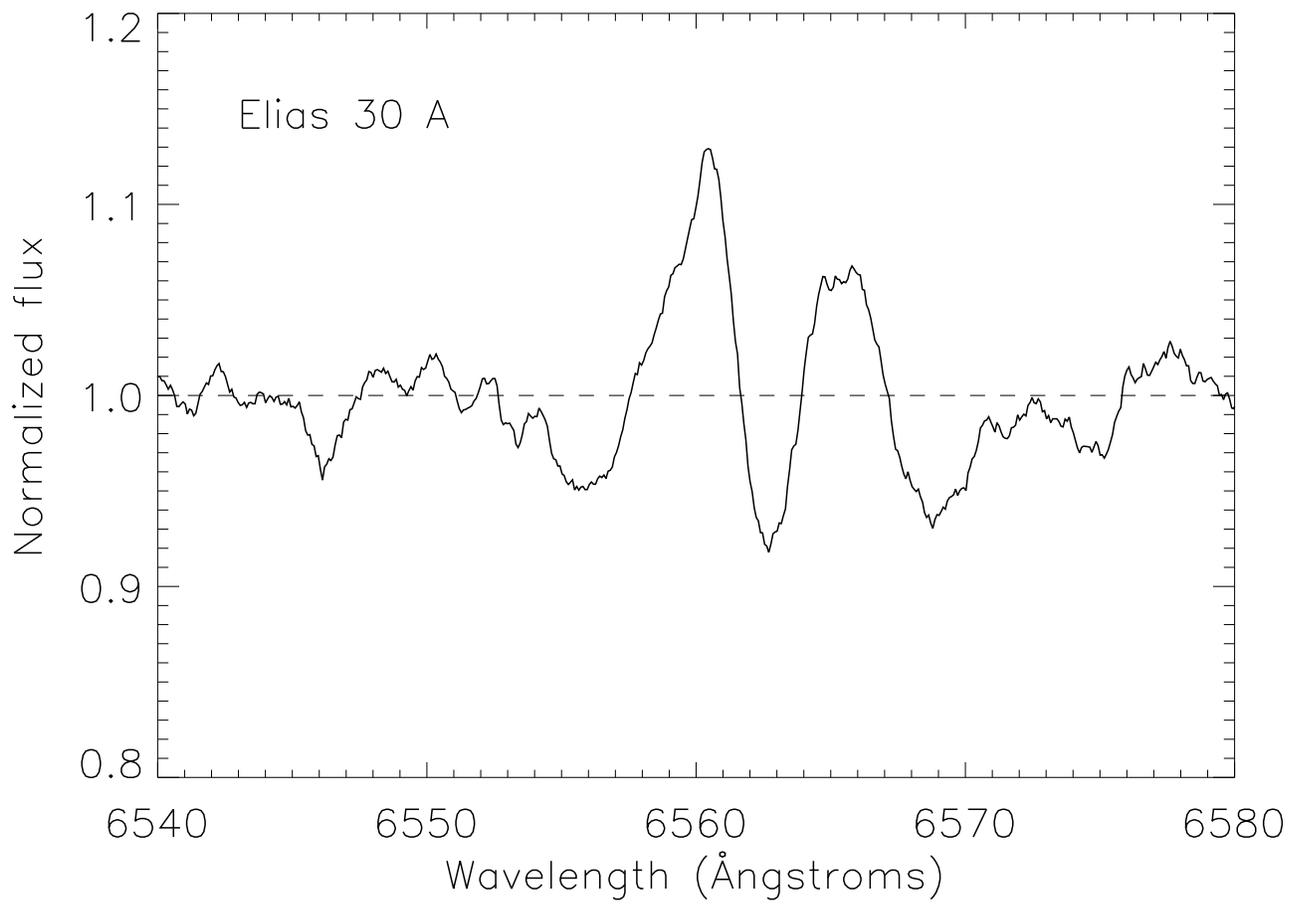

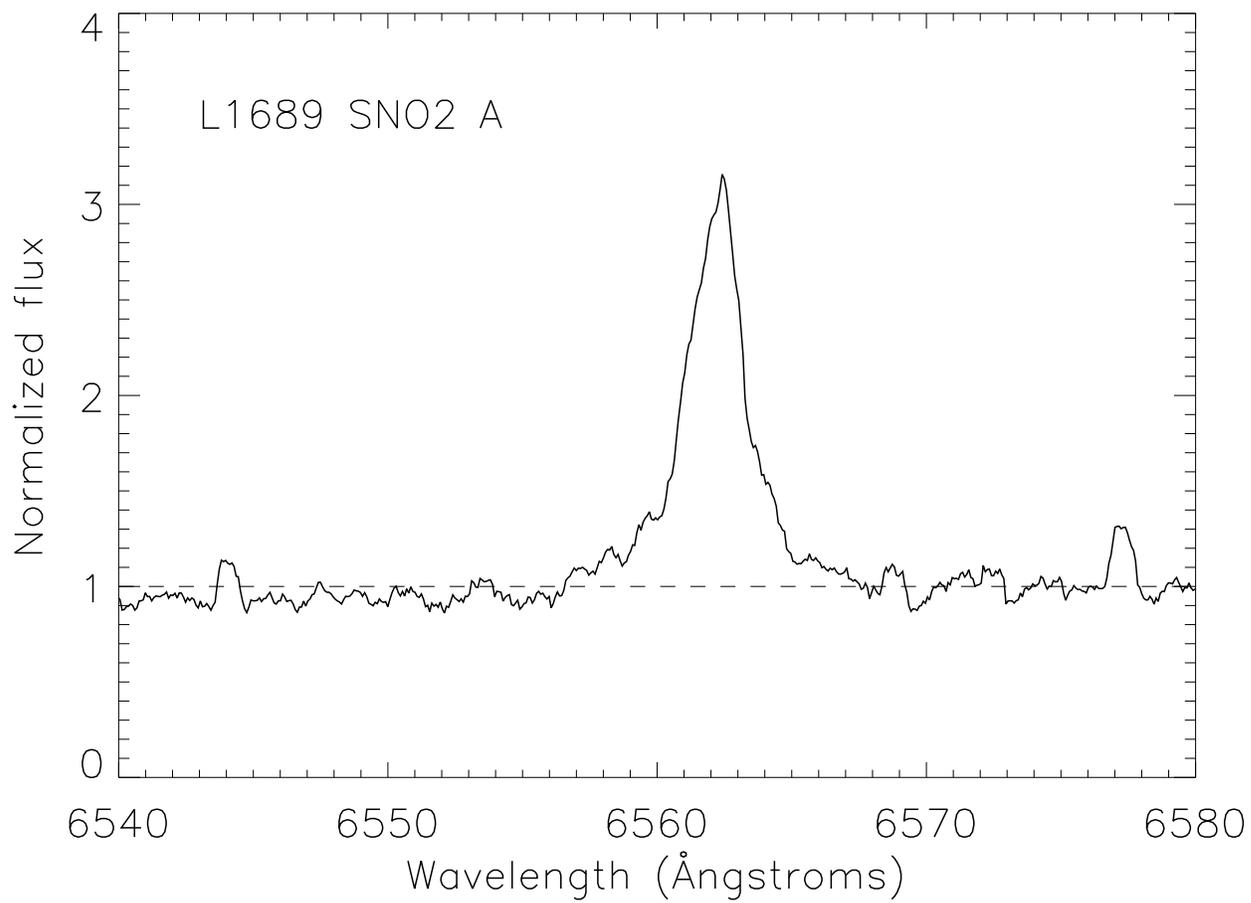

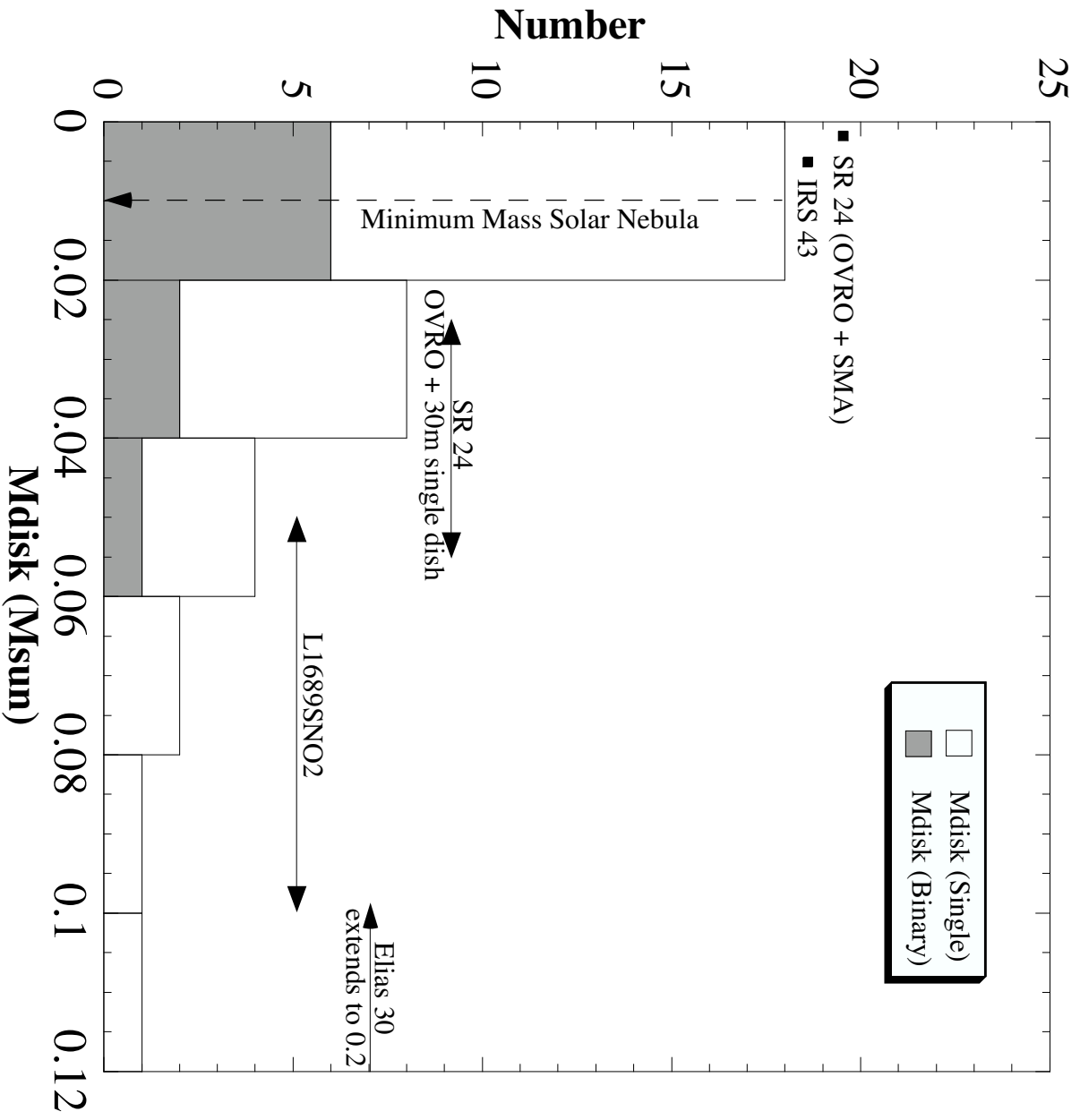

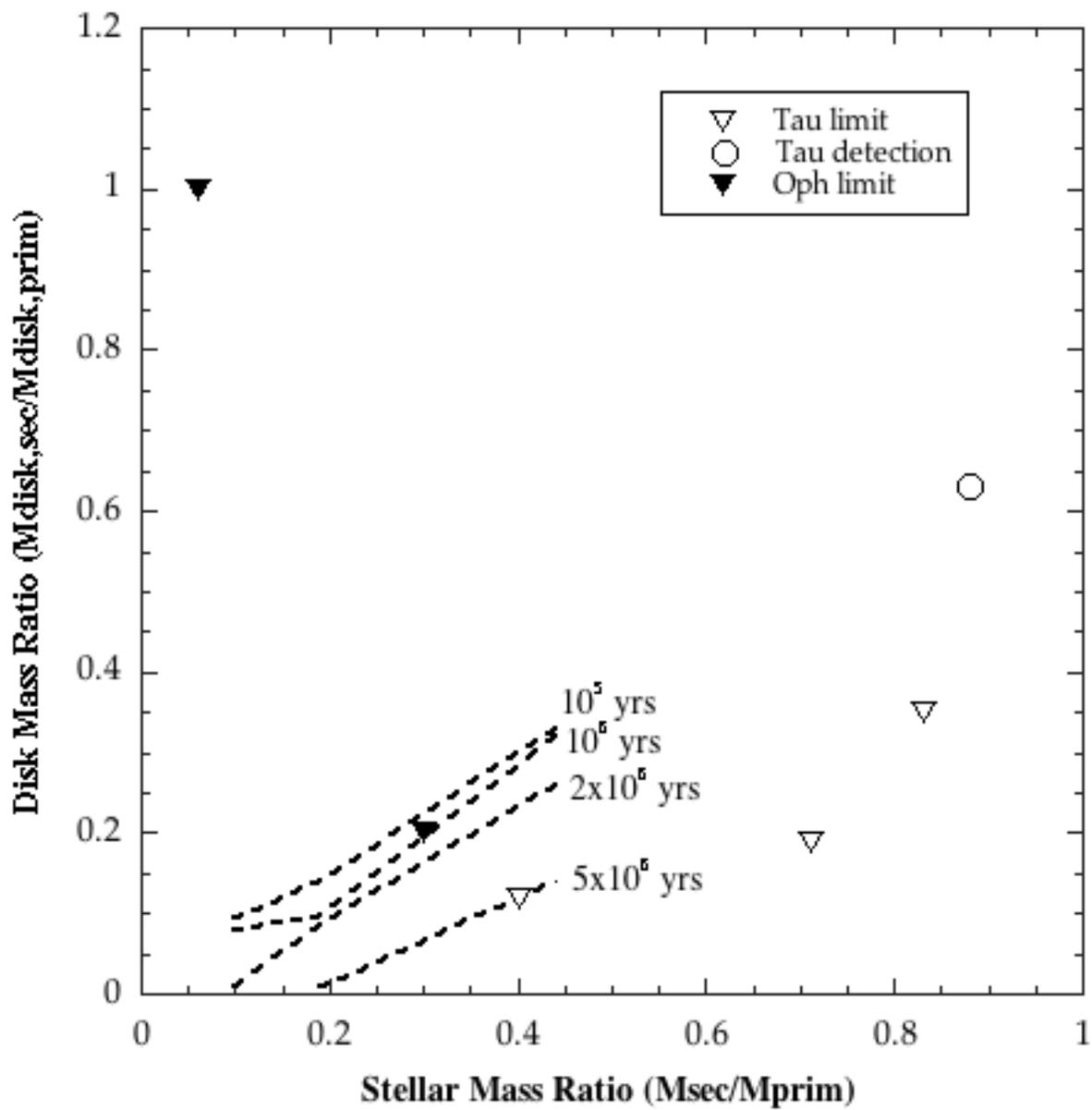